\documentclass[twocolumn, 10pt, journal,letterpaper]{IEEEtran}
\IEEEoverridecommandlockouts
\usepackage{algorithm}
\usepackage{graphicx}
\usepackage[noend]{algpseudocode}
\usepackage{url} 
\usepackage{etex} 
\usepackage{tikz}
\usetikzlibrary{shapes,arrows}
\usepackage{pstricks}
\usepackage{pst-node,pst-blur}
\usetikzlibrary{arrows}
\usepackage{amsfonts}
\usepackage{tabularx}
\usepackage{enumerate}
\usepackage{subfigure}
\usepackage{amssymb}
\usepackage{steinmetz}   
\usepackage{booktabs}
\usepackage{multirow}
\usepackage{epsfig}
\usepackage{epstopdf}
\usepackage{array}

\usepackage[noadjust]{cite}

\newtheorem{remark}{Remark}

\newtheorem{lemma}{Lemma}

\usepackage[cmex10]{amsmath}
\usepackage[cmex10]{mathtools}

\newcommand\abs[1]{\left\lvert#1\right\rvert}
\newcommand\norm[1]{\left\lVert#1\right\rVert}
\algdef{SE}[DOWHILE]{Do}{doWhile}{\algorithmicdo}[1]{\algorithmicwhile\ #1}%

\allowdisplaybreaks  

\makeatletter 
\def\@eqnnum{{\normalsize \normalcolor (\theequation)}} 
\makeatother

\begin{document}   
\title{Multi-Tag Backscattering to MIMO Reader: Channel Estimation and Throughput Fairness} 
\author{Deepak Mishra,~\IEEEmembership{Member,~IEEE,} and Erik G. Larsson,~\IEEEmembership{Fellow,~IEEE}
\thanks{D. Mishra is with the School of Electrical Engineering and Telecommunications at the University of New South Wales (UNSW) Sydney, NSW 2052, Australia (email: d.mishra@unsw.edu.au).}
\thanks{E. G. Larsson is with the Department of Electrical Engineering (ISY) at the Link\"oping University, Link\"oping 58183, Sweden (email: erik.g.larsson@liu.se).}
\thanks{This work was supported by  ELLIIT and Swedish Research Council (VR).}
\thanks{{A preliminary  version~\cite{SPAWC19} of this work was presented at the IEEE SPAWC, Cannes, France in July 2019.}}}
\maketitle

\begin{abstract}	
Green low power networking with the least requirement of dedicated radio resources is need of the hour which has led to the upsurge of backscatter communication (BSC) technology. However, this inherent potential of BSC is challenged by hardware constraints of the underlying tags. We address this timely concern by investigating the practical efficacy of multiple-input-multiple-output (MIMO) technology in overcoming the fundamental limitations of BSC. Specifically, we first introduce a novel least-squares based channel estimation (CE) protocol for multi-tag BSC settings that takes care of both the unintended ambient reflections and the inability of tags in performing estimation by themselves. {Then using it, a nontrivial low-complexity algorithm is proposed to obtain the optimal transceiver designs for the multiantenna reader to maximize the minimum value of the lower-bounded backscattered throughput among the single-antenna semi-passive tags.} Additional analytical insights on both individually and jointly-optimal precoding vector and detector matrix at the reader are provided by exploring the asymptotically-optimal transceiver designs.   Lastly detailed numerical investigation is carried out to validate the theoretical results and quantify the practically realizable throughput fairness. Specifically, more than seven-fold increase in the common-backscattered-throughput among tags as achieved by the proposed designs over the relevant benchmarks corroborates their practical significance. 
\end{abstract}


\section{Introduction}\label{sec:intro} 
Backscatter communication (BSC) technology, involving the low power tags as the information sources that do not require
costly and bulkier conventional  radio frequency (RF) chain components like local oscillators, mixers, and converters~\cite{Survey-AmbBSC}, has recently emerged as a very reliable solution for  enabling the sustainable pervasive computing demand in Internet of Things (IoT)~\cite{New-SP-Mag2}. Thus, with these resource-constrained tags relying on the reflection-based alternative modulations~\cite{New-SP-Mag1} using only passive elements like envelope detectors,  comparators, and impedance controllers, this BSC technology can also help in the ubiquitous deployment of low-cost wireless devices in 5G networks. 
Despite these potential merits, the widespread applicability of BSC is limited by the lower end-to-end communication range and lower achievable data rates~\cite{TSP19}. Therefore,  to overcome these challenges and thereby enable efficient BSC from multiple tags, there is a need  for developing novel channel estimation (CE) protocols and optimal transceiver (TRX) designs for multiantenna full-duplex reader to efficiently utilize its available resources.


\subsection{State-of-the-Art}\label{sec:rw}
BSC technology, not requiring any active RF chain components at the tags, can yield significant savings in terms of hardware cost and energy consumption. However, due to the unavailability of   dedicated radio transmission circuitry at tags, the reader has to perform two tasks: (a) downlink (DL) carrier transmission to tags for their excitation, and (b) uplink (UL) detection of the modulated backscattered signals~\cite{inv-BSC-MIMO} from the tags. Based on whether these two tasks are performed by the same entity or not, there are different BSC configurations: monostatic and bi-static~\cite{New-SP-Mag1,New-SP-Mag2}.   As the achievable beamforming and multiplexing gains of using antenna array at the reader are strongly influenced by the underlying  channel state information (CSI), CE quality becomes very crucial. Here we first present the prior art~\cite{RFID-TSP,WaveOpt-BSC,MultiAccessTag,CogBSC,TCOM-BSC,ICASSP-SRM} on the TRX designs for the multiantenna reader under the assumption that it has the perfect CSI availability.   Considering frequency-modulated continuous-wave based monostatic BSC system with multiantenna reader, whose one antenna was dedicated for transmission and remaining for  the reception of backscattered signals, authors in~\cite{RFID-TSP} focused on determining the number of active tags and their positions.  More recently, the transmit  waveform  and
receive  combiner design for wirelessly powered multi-tag BSC was studied in~\cite{WaveOpt-BSC} under the assumption of perfect CSI availability for BSC channels to investigate the tradeoff  between  harvested  energy  among tags  and  signal-to-interference-plus-noise ratio (SINR) at the reader during information decoding (ID). 
For the single antenna ambient BSC systems, a backscatter multiplicative multiple-access channel scheme was introduced in~\cite{MultiAccessTag}, whose  achievable  rate  region  was shown to be larger than that of the conventional  time-division multiple-access systems. This corroborates the practical utility of BSC in addressing the spectrum-scarcity constraints. With similar objective, cognitive ambient BSC was investigated in~\cite{CogBSC} to come up with beamforming-assisted energy detector and likelihood-ratio based detector  designs for the multiantenna reader having the  perfect
CSI availability. More recently, under perfect CSI availability assumption, the sum-backscattered-throughput maximization problem, jointly optimizing TRX design
at multiantenna reader and backscattering coefficient at single antenna tags, was studied in \cite{TCOM-BSC,ICASSP-SRM}.

Now we would like to highlight a striking similarity between BSC system~\cite{Survey-AmbBSC,TSP19,New-SP-Mag1,New-SP-Mag2,inv-BSC-MIMO,semi-passive,FD-General,FD-conf-new,BackFi,WaveOpt-BSC,TCOM-BSC,ICASSP-SRM,MultiAccessTag,CogBSC,RFID-TSP} and the wireless powered communication network (WPCN)~\cite{WPCN-MIMO-CTM,ZF-CTM}. The role of hybrid access point (HAP), providing downlink energy transfer in WPCN to power-up the RF energy harvesting (EH) users, is analogous to the reader's carrier transmission to the tags for exciting them. Thus, the proposed TRX designs can be applied for the common-throughput-maximization in WPCN with multiantenna HAP and multiple single-antenna EH users~\cite{WPCN-MIMO-CTM,ZF-CTM}. This application or scope of our work is important because the proposed optimal TRX design outperforms the suboptimal solutions of~\cite{WPCN-MIMO-CTM,ZF-CTM}, as shown via numerical results in Section~\ref{sec:res}. It may be recalled that since the EH devices have their own dedicated RF chain, so sophisticated TRX designs can be implemented by them also, which resource-constrained tags cannot do. Also recently, a couple of works~\cite{WPCN-BSC1,WPCN-BSC2}  investigated hybrid multi-tag BSC for WPCN, but with single antenna HAP (reader) and  perfect CSI availability, for maximizing the underlying sum-throughput performance.

Finally, let us shift focus to the works that did not consider perfect CSI assumption in BSC systems~\cite{BSC-WET-MIMO,CE-CL-BSC,Interference-BSC,SPAWC18, TSP19}. With tag-to-reader channel partially known at the reader, in~\cite{BSC-WET-MIMO} an estimate for the reader-to-tag channel was obtained by exploiting reciprocity to eventually derive the precoder design for optimizing wireless energy transfer to a tag when only one of the reader's antenna is available for ID. Authors in \cite{CE-CL-BSC} investigated blind CE algorithms for obtaining the estimates for :(a) absolute  values of the channel coefficients for RF source to tag link, and (b) the scaled product of forward and backward links in BSC. 
Alternatively, robust inference algorithms not requiring any CSI were proposed in~\cite{Interference-BSC} for detecting the backscattered information from multiple single-antenna sensors at a multiantenna reader. Very recently, a least-squares (LS) based channel estimator for the forward and backward channels between a multiantenna reader and single-antenna tag was proposed in~\cite{SPAWC18} to come up with an optimal energy allocation scheme  maximizing underlying BSC performance. Based on that a minimum-mean square-error  based channel estimator has also been designed in~\cite{TSP19} while investigating the optimal number of orthogonal pilots for CE in a single-tag BSC setting.  Lastly, it is also worth noting that the several existing CE algorithms that have been developed for the two-way AF relay channel modelling set-ups~\cite{Prod-Relay-CE-TSP,Prod-Relay-CE-TVT} cannot be used in BSC settings because the tags do not have the required radio resources like a relay for implementing the two-phased CE algorithms. In those CE algorithms, the first phase involved source-to-relay CE by relay, and then during the second phase, the cascaded source-to-relay-destination channel was estimated by destination using the first phase's CE outcome sent as a  feedback by the relay.

\subsection{Paper Organization and Notations Used}\label{sec:pap-org}
{We start with outlining the practical scope  and  key  contributions of this work in Section \ref{sec:motiv}. Then we present the adopted BSC system model and the proposed LS-based CE protocol  in Sections~\ref{sec:model} and~\ref{sec:prot}, respectively.} Thereafter, the joint TRX design problem at the multiantenna a reader for the common throughput maximization (CTM) among the tags is outlined in Section~\ref{sec:fair} along with the insights on individually-optimal designs. {Here, due to the adoption of linear TRX designs, we have  actually optimized a lower-bound on BSC throughput.} Next after detailing-out the asymptotically-optimal TRX designs in Section~\ref{sec:sol}, we present a novel solution methodology to yield an efficient near-optimal TRX design for the reader. Extensive numerical investigation is carried out in Section~\ref{sec:res}, followed by the concluding remarks summarized in Section~\ref{sec:concl}.

Throughout this paper, vectors and matrices are respectively denoted by boldface lowercase and capital letters. $\mathbf{A}^{\mathrm H}$, $\mathbf{A}^{\mathrm T}$, and $\mathbf{A}^{\mathrm *}$ respectively denote the Hermitian transpose, transpose, and conjugate of matrix $\mathbf{A}$. $\mathbf{0}_{n\times n},\mathbf{1}_{n\times n},$ and $\mathbf{I}_{n}$ respectively represent  $n\times n$ zero, all-ones, and identity matrices.  $[\mathbf{A}]_{i,j}$ stands for   $(i,j)$-th element of matrix $\mathbf{A}$ and $[\mathbf{a}]_{i}$ stands for  $i$-th element of vector $\mathbf{a}$. With $\mathrm{Tr}\left\lbrace\mathbf{A}\right\rbrace$  being the trace,  $\lVert\,\cdot\,\rVert$ and $\left|\,\cdot\,\right|$ respectively represent Frobenius norm of a complex matrix and absolute value of a complex scalar. 
The principal eigenvector corresponding to maximum eigenvalue $\lambda_{\max}\left\lbrace\mathbf{A}\right\rbrace$ of  $\mathbf{A}$ is represented by $\mathrm{v}_{\max}\left\lbrace\mathbf{A}\right\rbrace$. Operators $\odot$ and $\otimes$ are respectively used to denote the 
Hadamard and   Kronecker products of two matrices.  $\mathrm{Re}\{\mathrm{a}\}$
and $\mathrm{Im}\{\mathrm{a}\}$ are used for representing the real and imaginary parts of the complex quantity $\mathrm{a}$. Lastly, with $j=\sqrt{-1}$, $\mathbb{Z}, \mathbb{R},$ and  $\mathbb{C}$ respectively denoting the integer, real, and complex number sets, $\mathbb{C} \mathbb{N}\left(\boldsymbol{\mu},\mathbf{C}\right)$ denotes complex Gaussian distribution with mean $\boldsymbol{\mu}$ and covariance matrix  $\mathbf{C}$.
\section{Motivation and Significance}\label{sec:motiv}
Here we highlight the research gap addressed and the practical scope of this work, followed up the key contributions made so as to corroborate it's significance and utility.

\subsection{Novelty and Scope}\label{sec:novelty}\color{black}
From the above detailed literature review, we notice that the existing CE algorithms~\cite{BSC-WET-MIMO,CE-CL-BSC,Interference-BSC,SPAWC18, TSP19,Prod-Relay-CE-TSP,Prod-Relay-CE-TVT} for BSC are either dedicated for single tag set-ups~\cite{SPAWC18, TSP19}, or are based on some strong assumptions on tags' capabilities~\cite{Prod-Relay-CE-TSP,Prod-Relay-CE-TVT}, or do not focus on obtaining the actual complex values of the individual forward and backward channel coefficients~\cite{BSC-WET-MIMO,CE-CL-BSC,Interference-BSC}. Hence, a low-complexity CE algorithm for estimating the forward and backward channels in a multi-tag BSC system with multiantenna reader, while keeping in mind the tags' resource constraints along with the unintended ambient reflections (UAR) and no strong prior availability on CSI, is missing in the existing art. We try to fill  this important research gap by presenting a novel CE protocol that addresses all these timely practical requirements of multi-tag BSC with MIMO reader.  Furthermore, as fair allocation of the energy resources of the reader is a major concern among the semi-passive tags, we use these estimates for BSC channel for eventually designing the optimal TRX at the reader which can maximize the common (or the minimum) backscattered throughput among the tags. \textit{Hence, to the best of our knowledge this is the first investigation on LS-based CE protocol for multi-tag BSC while accounting for UAR and tags' limitations. Also, we propose novel asymptotically-optimal TRX designs that can maximize the efficacy of multiantenna reader in significantly enhancing the underlying fair throughput performance.}  {The key distinguishing features of our proposed CE protocol from the conventional ones designed for multiuser multiantenna communications~\cite{massive-MIMO} include: (a) no pilot transmission from the hardware-limited tags, (b) novel preamble design to estimate BSC channel gain coefficients in a multi-tag set-up, (c) obtaining the globally-optimal  LS estimators from the underlying product or cascaded channel estimates in a computationally-efficiently manner, and (d) effective detection and suppression of UAR to maximize the practical  efficacy of full-duplex multiantenna reader.}

{We consider the monostatic BSC configuration~\cite{inv-BSC-MIMO}, where both the carrier emitter and the backscattered signal reader  are the same entity. Further, we assume that the reader has multiple antennas working in the full-duplex mode~\cite{TSP19} to simultaneously serve multiple single-antenna  semi-passive tags~\cite{semi-passive}. This full-duplex reader operation, involving efficient suppression of the unmodulated carrier leakage~\cite{BackFi}, provides   array gains during DL carrier transmission and also enables spatial multiplexing of the backscattered signals in the UL. Also, in contrast to the full-duplex operation in conventional wireless systems~\cite{FD-General} involving modulated information signals, the unmodulated carrier leakage can be efficiently suppressed~\cite{FD-conf-new} in monostatic full-duplex BSC systems.  Exploiting this fact, several practical implementations~\cite{RFID-bookch,FD-Just,FD-conf-new,FD-jnl-new,BackFi} for full-duplex readers   in monostatic BSC have been realized either using circulator or directional coupler. So, for enabling full-duplex operation, reader includes a decoupler (or a circulator \cite{RFID-bookch}) which comprises of automatic gain control circuits and conventional phase locked loops~\cite{FD-Just} to  effectively suppress out the transmit carrier from the backscattered one at the receiver unit~\cite{FD-conf-new}. Here, exploiting the fact that the reader performs an unmodulated transmission, this  decoupler can easily suppress the self-jamming carrier, while isolating the transmitter and receiver units' paths, to eventually implement the full-duplex architecture for monostatic BSC settings~\cite{FD-jnl-new,uspat-unmod1}. Due to this adopted full-duplex architecture involving the decoupler to separate out transmit and receive unit of the reader sharing same antennas, the channel reciprocity holds for the adopted time division duplex (TDD) mode in this paper. It is worth noting that channel-reciprocity is a common assumption in the monostatic multiantenna BSC architectures~\cite{TSP19,Retro-BSC,WPCN-BSC-Access} and there are commercially available readers~\cite{Impinj}) based on it. Also, in future we would like to extend our proposed CE and TRX designs methodology  to investigate the bi-static and ambient BSC configurations where the forward and backward links are different and thus, are non-reciprocal.}

The proposed low-complexity CE  protocol involving optimal TRX designing for the monostatic BSC configuration  shifts the high cost and large form-factor constraints to the reader side, which eventually leads to the \textit{tag-size miniaturization} and cost reduction. Also, this CE protocol can be extended to address the demands of point-to-point BSC between a multiantenna tag and reader. Further, due to the mentioned similarity with WPCN, the  scope of the TRX deigns discussed here also includes precoder and detector designs for HAP in similar product channel settings~\cite{WPCN-MIMO-CTM,WPCN-BSC1,WPCN-BSC2,ZF-CTM,Prod-Relay-CE-TSP,Prod-Relay-CE-TVT,BSC-WET-MIMO,CE-CL-BSC,Interference-BSC,SPAWC18}. The key practical merit of the proposed CE protocol with optimal TRX design is that it does not require any prior information and needs extremely limited or minimal participation of the resource-constrained tags in obtaining the channel estimate as the global minimizer of the LS problem under phase ambiguity. The merits of these proposals include longer BSC range and throughput fairness maximization. Lastly, the novel analytical results provided in terms of asymptotically-optimal TRX or individually-optimal precoder and detector designs, shed nontrivial insights on BSC \textit{systems engineering} perspectives.

\subsection{Contributions of the Paper}\label{sec:contib} 
This work tries to address the timely demand of investigating the practical efficacy of utilizing MIMO technology at reader  to overcome  limitations of tags to eventually realize sustainable quality-of-service (QoS)-aware BSC. The five-fold key contribution  is discoursed below.
\begin{itemize}
	\item A novel CE protocol is introduced for the multi-tag monostatic BSC system with full-duplex reader that takes care of the practical constraints like unintended ambient reflections (UAR), preamble design for the tags, and non-convexity of the underlying LS estimation problem.
	\item To ensure fair allocation of multiantenna reader's resources, we  investigate the nontrivial problem of CTM among tags by optimally designing the TRX.
	\item Since CTM problem is NP-hard~\cite{P-NP}, we disclose new asymptotically optimal TRX designs along with  individually optimal precoding vector and detector matrix. 
	\item Using these valuable insights, we propose an efficient low complexity iterative algorithm that yields a near-optimal  TRX providing substantial gains over the benchmarks. 
	\item Lastly, selected numerical results validate the key analytical claims, provide optimal TRX design insights, and quantify the fair-throughput performance enhancement. 
\end{itemize}

Here we would like to mention that apart from the novel CE protocol designed specifically to address the practical challenges in multi-tag BSC settings, the proposed non-trivial TRX design takes into account the particular resource-constraints of the low-power BSC technology  while yielding significant fair-throughput enhancement over the existing benchmarks.

\section{System Description}\label{sec:model}
In this section we present the adopted network architecture, transmission protocol, BSC channel and tag models along with the formal definition for the problem tackled in this work.

\subsection{Network Architecture and Transmission Protocol}\label{sec:network} 
\begin{figure}[!t]
	\centering 
	\includegraphics[width=3.4in]{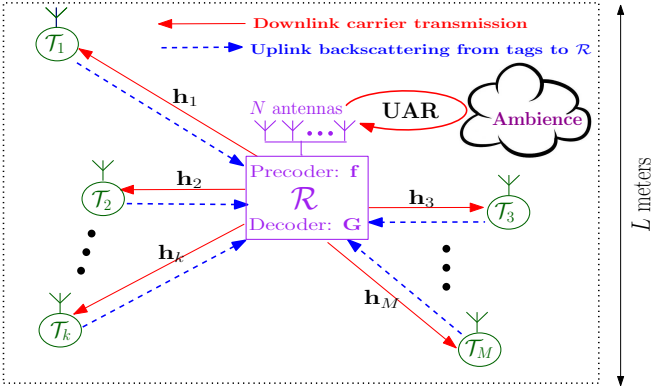}\caption{\small Adopted monostatic BSC model with a full-duplex $N$-antenna reader and $M$ semi-passive single-antenna tags, that exploits the proposed CE protocol  with optimal TRX design $\left(\mathbf{f},\mathbf{G}\right)$ for reader.}
	\label{fig:model} 
\end{figure}

We consider a multi-tag  MIMO monostatic BSC system comprising of $M$ single-antenna semi-passive tags, as denoted by $\mathcal{T}_k,$ $\forall k\in\mathcal{M}\triangleq \{1,2,\ldots,M\}$, and one multiantenna reader (cf. Fig.~\ref{fig:model}). This reader $\mathcal{R}$, having $N$ antennas, works in   full-duplex mode~\cite{FD-Just} and is responsible for simultaneously transmitting carrier signal to the tags and decoding the underlying backscattered signals. As shown in Fig.~\ref{fig:model}, the $M$ semi-passive tags are randomly deployed in a square field of length $L$ meters (m), following  Poisson point process, with $\mathcal{R}$ being at center.   

As mentioned earlier, due to the unavailability of the desired radio resources at the tags, the CE and corresponding TRX optimization operation are performed at   $\mathcal{R}$, which has the required radio and computational resources. Also, since during the backscattered data detection, called as information decoding phase (ID) in this work, $\mathcal{R}$ does not have the information on the scaled pilot matrix because the tags have added their data on top of the received carrier signal from $\mathcal{R}$, we need to consider \textit{two-phase transmission protocol} for each coherence interval of $\tau$ samples. Here during the first phase, called CE, the tags cooperate with the $\mathcal{R}$ and set their reflection coefficients to a value, $\zeta=\zeta_{\rm off}$~\cite{BSC-WET-MIMO,TSP19}, known to  $\mathcal{R}$ over the first $\tau_c\le\tau$ samples. This can be seen as a preamble of length $\tau_c$ samples  for the data transmission from the tags. Thus, with $\mathcal{R}$ transmitting pilot matrix $\mathbf{S}$ and having tags cooperation in terms of reflecting back the known complex baseband signal, this knowledge of scaled pilot matrix at $\mathcal{R}$ is utilized in the CE process for performing the matched-filtering operation on the received signal.  Details of the proposed CE process where this preamble designing~\cite{New-SP-Mag1} at the tags also involves suppression of the UAR are presented in Section~\ref{sec:prot}. Using this estimate obtained after CE phase, the second phase involves  ID  of  the unknown modulated backscattered signal from the tags over the remaining $\tau-\tau_c$ samples. This BSC channel estimate is also used for the TRX designing at $\mathcal{R}$ during ID phase.

\subsection{Adopted BSC Channel Model and Tag Models}\label{sec:tag}
{Due to the rich-backscattering environment in short range BSC settings with multiple tags and UAR, we assume that all the links suffer from the flat quasi-static Rayleigh block fading~\cite{TSP19,inv-BSC-MIMO,Amb_MSC-Model,Retro-BSC}.} Thus, the channel impulse response for each link remains constant during a coherence block of $\tau$ samples, and varies independently across different coherence intervals. So, with parameter $\beta_i$ representing the average channel power gain that incorporates both the fading gain and propagation loss over $\mathcal{T}_i$-to-$\mathcal{R}$ link, the corresponding wireless channel is denoted by an $N\times1$ vector $\mathbf{h}_i\sim\mathbb{C} \mathbb{N}\left(\textbf{0}_{N\times 1},\beta_i\,\mathbf{I}_N\right),\forall\,i\in\mathcal{M}$. This leads to the cascaded MIMO system as defined by the transmission chain $\mathcal{R}$-to-tags-to-$\mathcal{R}$. {Here it may be noted any fading model can be adopted as our CE protocol and TRX designing methodology are  generic in nature and do not depend on the underlying distribution of the channel fading gains. So, our proposals and analyses  can also be readily used for Rician or any other fading channels.}

We consider that the multiantenna $\mathcal{R}$ adopts linear beamforming and assigns a precoding vector $\mathbf{f}$ to the unit power carrier signal  $\mathrm{c}\in\mathbb{C}$ for the DL carrier transmission. As such, the complex baseband transmitted signal from  $\mathcal{R}$ is given by $\mathbf{x}_{\mathcal{R}}\triangleq\mathbf{f}\,\mathrm{c}\in\mathbb{C}^{N\times 1}$ with a total power budget $p_t$ constraint such that $\norm{\mathbf{f}}^2\le p_t$.   The resulting $M$ data symbols are simultaneously backscattered from the $M$ tags and are spatially separated at $\mathcal{R}$ with the aid of $M$ linear  decoding vectors $\mathbf{g}_1,\mathbf{g}_2,\ldots,\mathbf{g}_M\in\mathbb{C}^{N\times 1}$. It may be noted that the linear precoding and decoding architectures have been adopted due to their very-low computational complexity in implementation as required for BSC and also linear TRX designs are known to be asymptotically-optimal~\cite{massive-MIMO}, i.e., for the settings with $N\gg M.$ So, focusing on the excitation of tags by carrier transmission from $\mathcal{R}$, the baseband received signal $\left[\mathbf{y}_{\mathcal{T}}\right]_k\in\mathbb{C}$ at $\mathcal{T}_k$ can be mathematically expressed as
\begin{equation}\label{Eq:System}
\left[\mathbf{y}_{\mathcal{T}}\right]_k= \mathbf{h}_{k}^{\rm T}\,\mathbf{x}_{\mathcal{R}} + \left[\mathbf{w}_{\mathcal{T}}\right]_k=\mathbf{h}_{k}^{\rm T}\,\mathbf{f}\,\mathrm{c} + \left[\mathbf{w}_{\mathcal{T}}\right]_k \quad\forall k\in\mathcal{M},
\end{equation}\color{black}
where   $\mathbf{w}_{\mathcal{T}}\in\mathbb{C}^{N\times 1}$ denotes the zero-mean Additive White Gaussian Noise (AWGN) with variance $\sigma_{\mathrm{w}_\mathcal{T}}^2$. Thus, the total received power at each $\mathcal{T}_k$ as available for  backscattering is 
\begin{equation}\label{eq:RP}
q_{k} \triangleq \left|\mathbf{h}_{k}^{\rm T}\mathbf{f}\right|^2 + \sigma_{\mathrm{w}_\mathcal{T}}^2,\quad\forall k\in\mathcal{M}.
\end{equation} 
 
On the excitation with the above power $q_{k}$, each $\mathcal{T}_k$ modulates the carrier received from $\mathcal{R}$ via a complex baseband signal denoted by $\left[\mathbf{x}_{\mathcal{T}}\right]_k\triangleq A_k-\zeta_{(k)}$~\cite{New-SP-Mag1}. Here, the load-independent constant $A_k$ is related to the antenna structure and the load-controlled reflection coefficient $\zeta_{(k)}\in\{\zeta_1,\zeta_2,\ldots,\zeta_V\}$ switches between $V$ distinct values to implement the desired tag modulation~\cite{Survey-AmbBSC}. Since we assume semi-passive tags, entire excitation power $q_k$ at $\mathcal{T}_k$ is  used for BSC. 

\subsection{Problem Definition}\label{sec:prob}
We aim to maximize the minimum backscattered throughput among the tags by optimizing the TRX design at $\mathcal{R}$ for the ID phase of $\left(\tau-\tau_c\right)$ samples using the  estimate for BSC channel as obtained after the CE phase. This objective leads to the  CTM among the multiple tags during BSC with a full duplex multiantenna reader. The  considered optimization model helps in ensuring the fair allocation of the reader's resources among the tags. This is important especially in IoT applications, where each tag has to report critical the sensed information back to $\mathcal{R}$ and periodic sensor readings from each tag are  important, regardless of the strengths of the individual links. 
 
{Now before deriving  backscattered throughput $\mathrm{R}_k$ for each $\mathcal{T}_k$, we recall that the practical backscattering due to the received noise signal $\left[\mathbf{w}_{\mathcal{T}}\right]_k$ at  the tags is negligible~\cite{RFID-TSP,WaveOpt-BSC,MultiAccessTag,CogBSC,TCOM-BSC,ICASSP-SRM,WPCN-MIMO-CTM,ZF-CTM,WPCN-BSC1,WPCN-BSC2,BSC-WET-MIMO,CE-CL-BSC,Interference-BSC,SPAWC18} in comparison to the corresponding received carrier reflection strength due to the signal $\mathbf{h}_{k}^{\rm T}\mathbf{f}\,\mathrm{c}$. Therefore, using \eqref{Eq:System} and ignoring the excitation of tags due to noise power $\sigma_{\mathrm{w}_\mathcal{T}}^2$, the received} {signal vector $\mathbf{y}_{\mathcal{R}}$ available for ID at $\mathcal{R}$  can be practically-approximated to as shown below}
\begin{align}\label{eq:yi}
\mathbf{y}_{\mathcal{R}} \triangleq&\,\sum_{k\in\mathcal{M}}\mathbf{h}_{k} \left[\mathbf{y}_{\mathcal{T}}\right]_k\,\left[\mathbf{x}_{\mathcal{T}}\right]_k + \mathbf{w}_{\mathcal{R}}\nonumber\\
\approx&\,\sum_{k\in\mathcal{M}}\mathbf{h}_{k}\,\mathbf{h}_{k}^{\rm T}\,\mathbf{f}\,\left[\mathbf{x}_{\mathcal{T}}\right]_k\,\mathrm{c} + \mathbf{w}_{\mathcal{R}},
\end{align}
where $\mathbf{w}_{\mathcal{R}}$ is the received noise at $\mathcal{R}$ with the variance for its entries being $\sigma_{\mathrm{w}_\mathcal{R}}^2$. We assume that $\mathbf{w}_{\mathcal{R}}$ includes  contributions from both   AWGN and unsuppressed UAR at $\mathcal{R}$ during ID.

With linear receive beamforming adopted at $\mathcal{R}$, this received signal $\mathbf{y}_{\mathcal{R}}$ is  multiplied with the  detection matrix ${\mathbf{G}}\triangleq\left[{\mathbf{g}}_1\;\;{\mathbf{g}}_2\;\;{\mathbf{g}}_3\;\ldots\;{\mathbf{g}}_M\right]\in\mathbb{C}^{N\times M}$ to obtain the decoded information vector $\widehat{\mathbf{y}}_{\mathcal{R}} \triangleq\mathbf{G}^{\rm H}\,\mathbf{y}_{\mathcal{R}}\in\mathbb{C}^{M\times1}$, whose each entry corresponding to each tag can be approximated to as
\begin{align}\label{eq:yd}
\left[\widehat{\mathbf{y}}_{\mathcal{R}}\right]_k\approx&\, \mathbf{g}_{k}^{\rm H}\,\mathbf{h}_{k}\,\mathbf{h}_{k}^{\rm T}\,\mathbf{f} \left[\mathbf{x}_{\mathcal{T}}\right]_k \mathrm{c} +\sum_{i=1,i\neq k}^{M}\mathbf{g}_{k}^{\rm H}\,\mathbf{h}_{i}\,\mathbf{h}_{i}^{\rm T}\,\mathbf{f} \left[\mathbf{x}_{\mathcal{T}}\right]_i \mathrm{c}\nonumber\\
&\, + \mathbf{g}_{k}^{\rm H}\,\mathbf{w}_{\mathcal{R}},\quad\forall k\in\mathcal{M}.
\end{align}

{We adopted linear TRX design to consider low-complexity signal processing and low-cost hardware implementation requirements of BSC settings having resource-constrained tags. However, with $N\gg M$, these linear precoder and detector designs tend to be nearly-optimal~\cite{massive-MIMO}. Therefore,  using approximation~\eqref{eq:yd} and considering linear TRX design, the resulting SINR available for decoding $\left[\mathbf{x}_{\mathcal{T}}\right]_k$ at $\mathcal{R}$ is lower-bounded by $\gamma_k$, which is defined as~\cite{Liu:2013:ABW,Retro-BSC,BSC-Relay,WET-Enabled,Amb_MSC-Model,Hybrid-BSC}}
\begin{align}\label{eq:SINR}
\gamma_k \triangleq \frac{q_{k} \left|\mathbf{g}_{k}^{\rm H}\,\mathbf{h}_{k}\right|^2}{ \sum\limits_{i\in\mathcal{M}_k}q_{i}\left|\mathbf{g}_{k}^{\rm H}\,\mathbf{h}_{i}\right|^2 + \overline{\sigma}_{\mathrm{w}_{\mathcal{R}}}^2\norm{\mathbf{g}_{k}}^2},
\end{align}
{where $\mathcal{M}_k\triangleq\mathcal{M}\setminus{\{k\}}=\{1,2,\ldots,k-1,k+1,k+2,\ldots,M\}$ and with $\overline{a}\triangleq\mathbb{E}\left\lbrace\abs{\left[\mathbf{x}_{\mathcal{T}}\right]_k}\right\rbrace,\forall k\in\mathcal{M},$ denoting the average amplitude of the tags' modulation during  ID phase~\cite{Bistatic-BCS,TSP19}, we assume that the normalized noise power is $\overline{\sigma}_{\mathrm{w}_{\mathcal{R}}}^2\triangleq\frac{\sigma_{\mathrm{w}_{\mathcal{R}}}^2}{\overline{a}^2}$. Since the instantaneous value of $\mathcal{T}_k$'s random message signal $\left[\mathbf{x}_{\mathcal{T}}\right]_k$ is unknown at $\mathcal{R}$ and has to be actually decoded while adopting the proposed CE and TRX designs, we consider it's mean value $\overline{a}$, which can be pre-defined and known at $\mathcal{R}$, while deriving the performance metric to be maximized.} Finally, using SINR expression in \eqref{eq:SINR}, the effective backscattered throughput for $\mathcal{T}_k,\forall k\in\mathcal{M},$ can be represented  as below 
\begin{eqnarray}\label{eq:Rk}
\mathrm{R}_k\left(\mathbf{f},\mathbf{G}\right)=\left(\frac{\tau-\tau_c}{\tau}\right)\log_2\left(1+\gamma_k\right).
\end{eqnarray}
{Above throughput expression which depends on the quality of the underlying received SINR has been widely adopted in BSC literature  for evaluating the performance over AWGN channels with perfect CSI availability and is similar to the performance metrics defined by~\cite[eq. (7)]{TCOM-BSC}, \cite[eq. (12)]{Amb_MSC-Model}, \cite[eq. (3)]{BSC-Relay}, and~\cite[eq. (6)]{WET-Enabled}.} Here, since maximizing the minimum $\mathrm{R}_k$ involves TRX designing based on the estimated BSC channels, we next present the proposed CE protocol, followed up  by the solution methodology for CTM.

\section{Multi-Tag Channel Estimation Protocol}\label{sec:prot} 
Here we first start with discoursing the different phases in the CE protocol, followed by the  LS optimization formulation and  corresponding LS estimators (LSE) for multi-tag BSC channels. This proposed CE process involves UAR suppression and yields the LSE in semi-closed form. {Before initiating the regular CE and BSC, $\mathcal{R}$ first identifies the interested tags in the neighbourhood by sending them a triggering signal to get them wake up from their respective sleep-states and get ready for backscattering. The tags within the read range, send acknowledgement signals to $\mathcal{R}$. Basically, $\mathcal{R}$ asks tags to backscatter their respective Electronic-Product-Code (EPC) and the latter respond back with their EPC. Thereafter, on detecting the $M$ tags of interest, $\mathcal{R}$ sends a clearance-to-send signal to them following similar procedure as in the EPC Global Class-1 Generation-2 (C1G2) protocol \cite{RFID-bookch}. Reacting to it, the tags get ready for the regular CE and backscattering phases to follow, which are discussed in detail next.} 
\subsection{Two-Phase CE Protocol for Multi-Tag BSC}\label{sec:Two-Phase-CE}
\begin{figure}[!t]
	\centering 
	\includegraphics[width=3.48in]{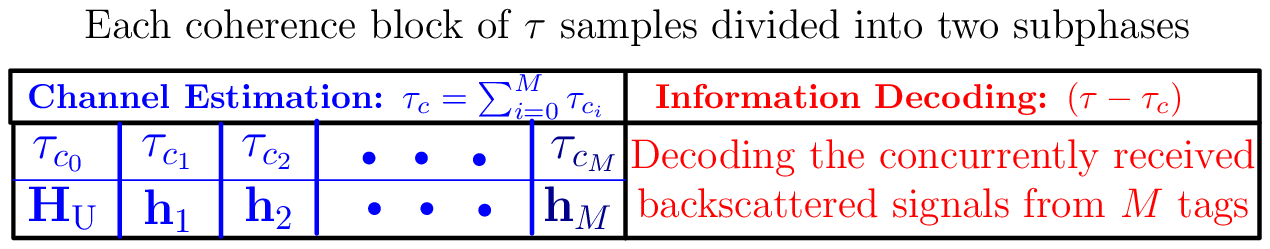} 
	\caption{\small Proposed two-phase (CE + ID) transmission protocol for the multi-tag BSC with full-duplex MIMO reader.}
	\label{fig:protocol} 
\end{figure}

We present a novel multi-tag backscattering protocol which  involves estimation of the channel vectors $\mathbf{h}_k,\,\forall\,k\in\mathcal{M}$ for the $M$ tags, where $N$ orthogonal pilots are used for CE, one from each antenna at $\mathcal{R}$.  Assuming channel reciprocity~\cite{TSP19,Retro-BSC,WPCN-BSC-Access}, the cascaded UL-DL channel coefficients are first estimated   during the CE phase  from backscattered  pilot signals,    isotropically transmitted from $\mathcal{R}$. Thereafter, the proposed methodology is adopted to suppress UAR and isolate the UL-DL coefficients. So,  each coherence interval of $\tau$ samples has two phases: (i) the CE phase involving the isotropic $N$ orthogonal pilot signals transmission, and (ii) the ID phase involving carrier transmission to tags and backscattered signals detection at $\mathcal{R}$ using the optimal TRX design based on the LSE for the backscattered channels obtained in the
first phase. The corresponding pictorial description for this protocol is provided in Fig.~\ref{fig:protocol}.  Here, noting the timing requirements for a practical tag protocol, as defined for the EPCglobal Class 1 Generation 2 standard (ISO 18000-6C), are to be longer than $1.56\times10^{-6}$ seconds (s)~\cite{RFID-bookch}, we assume that this length $\ell$ of each sample in time units (seconds) is more than $2\times10^{-6}$ s~\cite{BSC-WET-MIMO}.  Below we discuss in detail the different subphases involved in the proposed CE protocol.  
\begin{itemize}
	\item \textit{Phase I: Channel Estimation (CE) $\rightarrow$} It can be divided into the following two subphases.
	\begin{enumerate}[(a)]
		\item \textit{UAR Estimation and Suppression:} 	During this first sub-phase of $1\le \tau_{c_0}\triangleq\frac{\tau_c}{M+1}\le \tau$ samples duration,  $\mathcal{R}$ transmits $N$ orthogonal pilots, each of length $\tau_{c_0}$ samples, from each of the $N$ antennas and all the tags $\mathcal{T}_k,\,\forall k\in\mathcal{M},$ set their refection coefficients to $\zeta_{\rm off}$. This underlying $\zeta_{\rm off}$ value is set such as to ensure that the resulting $a_0\triangleq\abs{\mathrm{x}_{\rm off}}=\abs{A-\zeta_{\rm off}}\ll1$ (theoretically, $a_0=0$). In other words, $\mathbf{x}_{\mathcal{T}}=\mathrm{x}_{\rm off}\,\mathbf{1}_{M\times1}$ during this sub-phase of CE phase. As discussed in Section~\ref{sec:network}, these $N$ orthogonal pilots, as represented by pilot signal matrix $\mathbf{S}\in\mathbb{C}^{N\times \tau_{c_0}}$, satisfy $\mathbf{S}\,\mathbf{S}^{\mathrm H}=\frac{p_t}{N}\,\tau_{c_0}\,\mathbf{I}_N$. This phase helps in estimating the UAR as received at the $N$ antennas from the ambient sources in environment (cf. Fig.~\ref{fig:model}). Also, we assume that $\tau_{c_0}=N$, which in time units is $\tau_{c_k}=N\ell$ seconds. This estimate for the UAR is then  removed from the signal received in the next sub-phases to proceed with actual CE for the $M$ $\mathcal{R}$-to-tags BSC links. {Key property exploited here is the practical fact that like BSC channels, UAR also remain constant over a coherence block~\cite{BackFi,uspat-unmod1} and  fast changing orthogonal pilots can be used for  their effective estimation.}
		\item \textit{Multi-tag BSC CE:} During this second CE sub-phase of $\sum_{k=1}^{M}\tau_{c_k}\le \tau$ samples duration, $M$ BSC channel vectors $\mathbf{h}_k,\,\forall k\in\mathcal{M},$ are estimated, on a tag-by-tag basis. To be specific, during the $k$ sub-phase of this second phase, only the $k$th tag $\mathcal{T}_k$ is active with its coefficient set as $\zeta_{\rm on}$ over $\tau_{c_k}$ samples duration that ensures the resulting $a_1\triangleq\abs{\mathrm{x}_{\rm on}}=\abs{A-\zeta_{\rm on}}$. Ideally, this should be unity, i.e., $a_1=1$. Whereas all other tags are in inactive mode setting their coefficients as $\zeta_{\rm off}$. So, during $k$th sub-phase, $\left[\mathbf{x}_{\mathcal{T}}\right]_k=\mathrm{x}_{\rm on}$ and $\left[\mathbf{x}_{\mathcal{T}}\right]_i=\mathrm{x}_{\rm off},\,\forall i\in\mathcal{M}_k$. Let us define below the combined tags' modulation matrix $\boldsymbol{\mathcal{A}}\in\mathbb{C}^{M\times M}$ during CE phase, which in other words defines the preamble design for the tags, with its each column's entries  representing their respective baseband signals for CE,	
		\begin{equation}
		[\boldsymbol{\mathcal{A}}]_{im}\triangleq\begin{cases}
		a_1 \approx1, & \text{$i=m$}\\
		a_0 \approx0, & \text{$i\neq m$}
		\end{cases},\qquad \forall\,i,m\in\mathcal{M}.
		\end{equation}
		So, for $k$th column only $k$th tag $\mathcal{T}_k$ is in the full backscattering or reflection mode~\cite{MAP-BSC} (termed as `$\mathrm{on}$' or active state) providing maximum reflection strength backscattered to $\mathcal{R}$. Whereas all other tags remain in the silent  (non-backscattering) or `$\mathrm{off}$' state~\cite{BackFi}. Thus, with this cooperation of the tags, we can obtain LSE $\widehat{\mathbf{h}}_k$ for each of the $M$ BSC channels $\mathbf{h}_k,\,\forall k\in\mathcal{M},$ one-by-one, over a total duration of $\tau_c\triangleq\sum_{k=0}^{M}\tau_{c_k}$ samples. This cooperation of tags in phase 1 is analogous to the \textit{preamble designing} in conventional communications where the initial part of data transmission phase comprises of control signals sent for smooth identification or separating-out of simultaneously received signals.
	\end{enumerate}
	 
	\item \textit{Phase II: Information Decoding (ID) $\rightarrow$} During this last phase of the CE protocol, $\mathcal{R}$ simultaneously transmits carrier signal to the $M$ tags using the precoding vector $\mathbf{f}$ and decodes the resulting backscattered signals from $M$ tags using the proposed detector design over the remaining $\left(\tau-\tau_c\right)$ samples duration. As mentioned earlier, these precoding and decoding vectors are respectively denoted by $\mathbf{f}$ and $\mathbf{g}_k,\,\forall\,k\in\mathcal{M}$, and are decided based on the LSE $\widehat{\mathbf{h}}_k,\,\forall k\in\mathcal{M}$, as obtained after the CE phase after UAR suppression.
\end{itemize}

Next we derive LSE for  $M$ BSC channels in multi-tag setup.

\subsection{LS-Based Estimator for the Multi-tag BSC Channels}
In this part, we start with the discussion on UAR estimation and suppression. This is followed up the derivation of the LS-based estimates for the BSC channels in semi-closed form. 

 
\subsubsection{Suppression of UAR}\label{sec:UAR} 
{The unintended ambient reflections (or UAR) of the  carrier signal transmitted by $\mathcal{R}$ can be effectively cancelled by exploiting the known information about the carrier transmission at the directional coupler of $\mathcal{R}$~\cite{uspat-unmod1,uspat-unmod2}.  Here we notice that the UAR contain both the structural and unintended environmental reflections along with the internal leakage~\cite{uspat-unmod1}. These reflections to $\mathcal{R}$ from the unintended directions, include the ones from non-active tags being in $\mathrm{off}$ state and the other environmental sources like objects in the vicinity of $\mathcal{R}$ or due to $\mathcal{R}$'s imperfect self-cancellations. So, in contrast to the modulated reflections from the tags due to the underlying backscattering operation, these UAR are undesired and form an environmental distortion whose adverse effects can be overcome by additional CE and UAR suppression efforts by the reader. Later, we have numerically investigated the impact of practical CE errors in UAR suppression via Figs.~\ref{fig:CE} and \ref{fig:comp}.} Next, following the CE protocol details from Section~\ref{sec:Two-Phase-CE} and denoting all these UAR due to the pilot signal $\mathbf{S}\in\mathbb{C}^{N\times N}$ transmission during the first sub-phase (i.e., phase (1a)) of duration $\tau_{c_0}$ of CE phase  by the matrix $\mathbf{H}_{\rm U}\in\mathbb{C}^{N\times N}$, the resulting backscattered or received signal matrix $\mathbf{Y}_0\in\mathbb{C}^{N\times N}$ at $\mathcal{R}$ can be written as
\begin{equation}\label{eq:Rx-Y1}
\mathbf{Y}_0=\left(\mathbf{H}_{\rm U}+\textstyle\sum_{k\in\mathcal{M}}a_0\,\mathbf{h}_k\, \mathbf{h}_k^{\rm T}\right)\mathbf{S}+\mathbf{W}_0,
\end{equation}
where $\mathbf{W}_0\in\mathbb{C}^{N\times N}$ is the received AWGN at $\mathcal{R}$ during phase (1a) with the variance for its entries being $\sigma_{\mathrm{w}_0}^2$. Here, it may be recalled that during this sub-phase (1a), all the tags remain in the $\mathrm{off}$ state. Next, with $\mathbf{S}^\dagger\triangleq\mathbf{S}^{\mathrm H}\left(\mathbf{S}\, \mathbf{S}^{\mathrm H}\right)^{-1}$ denoting the pseudo-inverse~\cite{kay1993fundamentals} of the pilot  matrix $\mathbf{S}$ and $\widetilde{\mathbf{H}}_{\rm U}\triangleq\sum_{m\in\mathcal{M}}a_0\,\mathbf{h}_m\, \mathbf{h}_m^{\rm T}+\frac{N}{p_t\tau_{c_0}}\,\mathbf{W}_0\,\mathbf{S}^{\mathrm H}$ representing the underlying LS error in UAR estimation, the desired estimate $\widehat{\mathbf{H}}_{\rm U}=\mathbf{Y}_0\,\mathbf{S}^\dagger$ for  UAR $\mathbf{H}_{\rm U}$, as obtained using \eqref{eq:Rx-Y1}, is given by 
\begin{equation}\label{eq:LSE-U}
\widehat{\mathbf{H}}_{\rm U}=\mathbf{H}_{\rm U}+\sum_{m\in\mathcal{M}}a_0\,\mathbf{h}_m\, \mathbf{h}_m^{\rm T}+\frac{N\,\mathbf{W}_0\,\mathbf{S}^{\mathrm H}}{p_t\tau_{c_0}}=\mathbf{H}_{\rm U}+\widetilde{\mathbf{H}}_{\rm U}.
\end{equation} 


With the LSE of UAR given by $\widehat{\mathbf{H}}_{\rm U}$ in \eqref{eq:LSE-U}, we next focus on obtaining the estimates for  BSC channels. Following Section~\ref{sec:Two-Phase-CE} and with $\boldsymbol{\mathcal{A}}$ denoting the preamble matrix for the tags during this sub-phase of duration $\sum_{k=1}^{M}\tau_{c_k}$, the received signal matrix at $\mathcal{R}$ during phase (1b) of CE is
\begin{equation}
\mathbf{Y}_1= \left(\boldsymbol{\mathcal{A}}\otimes\mathbf{S}\right)\,\boldsymbol{\mathcal{H}}+\left(\mathbf{1}_{M\times1}\otimes\mathbf{H}_{\rm U}\right)\mathbf{S}+\mathbf{W}_1,
\end{equation}
where  $\boldsymbol{\mathcal{H}}\triangleq\left[\mathbf{h}_1 \mathbf{h}_1^{\rm T}\;\;\mathbf{h}_2 \mathbf{h}_2^{\rm T}\;\;\mathbf{h}_3 \mathbf{h}_3^{\rm T}\;\ldots\;\mathbf{h}_M \mathbf{h}_M^{\rm T}\right]^{\rm T}\in\mathbb{C}^{N M\times N},$  $\mathbf{W}_1\in\mathbb{C}^{N M\times N}$ is the AWGN matrix with  zero-mean independent and identically distributed entries having variance $\sigma^2_{\rm {w_1}}$. Here, it may be recalled that each $\mathbf{h}_k,\forall k\in\mathcal{M},$ and $\mathbf{H}_{\rm U}$ during each coherence block duration. Now before using this  received signal matrix $\mathbf{Y}_1\in\mathbb{C}^{NM\times N}$ for estimation of the $M$ BSC channels, we remove the UAR from it by subtracting its estimate $\widehat{\mathbf{H}}_{\rm U}$ as defined in \eqref{eq:LSE-U}.  Hence, the effective  received signal  $\mathbf{Y}\in\mathbb{C}^{NM\times N}$ at $\mathcal{R}$ during  phase (1b) of CE after UAR suppression is given by
\begin{eqnarray}\label{eq:rxR}
\mathbf{Y}= \mathbf{Y}_1-\left(\mathbf{1}_{M\times1}\otimes\widehat{\mathbf{H}}_{\rm U}\right)\mathbf{S}= \left(\boldsymbol{\mathcal{A}}\otimes\mathbf{S}\right)\,\boldsymbol{\mathcal{H}}+\widetilde{\mathbf{W}}, 
\end{eqnarray}
where  $\widetilde{\mathbf{W}}\triangleq\mathbf{1}_{M\times1}\otimes\left(\sum_{m\in\mathcal{M}}a_0\,\mathbf{h}_m\, \mathbf{h}_m^{\rm T}\,\mathbf{S}+\mathbf{W}_0\right)+\mathbf{W}_1$ is the effective noise term.

We next formulate the problem of LS estimation of BSC channels based on  this received signal  $\mathbf{Y}$ and this CE does not require any prior on statistics for $\boldsymbol{\mathcal{H}},\widehat{\mathbf{H}}_{\rm U},\mathbf{W}_0$ or $\mathbf{W}_1$.

\subsubsection{Least-Squares Optimization Formulation}\label{sec:LS-Opt}
The optimal LSE for the considered BSC channels between $M$ single-antenna tags and a  full-duplex monostatic MIMO reader can be obtained by solving the following constrained LS problem $ \mathcal{O}_{\rm L}$ in $ \mathbf{h}_1,\mathbf{h}_2,\mathbf{h}_3,\ldots,\mathbf{h}_M$,
\begin{align}\nonumber
\mathcal{O}_{\rm L}:\,& \underset{ \mathbf{h}_1,\mathbf{h}_2,\ldots,\mathbf{h}_M}{\text{argmin}}\;\mathcal{E}\triangleq
\norm{\mathbf{Y}-\left(\boldsymbol{\mathcal{A}}\otimes\mathbf{S}\right)\,\boldsymbol{\mathcal{H}}}^2,\\
&\text{s. t.}\quad({\rm C0}):\boldsymbol{\mathcal{H}}=\left[\mathbf{h}_1 \mathbf{h}_1^{\rm T}\;\;\mathbf{h}_2 \mathbf{h}_2^{\rm T}\;\ldots\;\mathbf{h}_M \mathbf{h}_M^{\rm T}\right]^{\rm T}.
\end{align}  
It is worth noting that $\mathcal{O}_{\mathrm L}$ is nonconvex in general. However, in the following we characterize all the possible candidates for the optimal solution of $\mathcal{O}_{\mathrm L}$ by setting the derivative of the objective $\mathcal{E}$ with respect to $\mathbf{h}_k,\forall k\in\mathcal{M},$ equal to zero and solving those system of equations to eventually obtain the optimal LSE. In this context, let us first rewrite  $\mathcal{E}$ in the following simplified form
\begin{align}\label{eq:tr-obj}
\mathcal{E}&=\mathrm{Tr}\left\lbrace\left(\mathbf{Y}-\left(\boldsymbol{\mathcal{A}}\otimes\mathbf{S}\right) \boldsymbol{\mathcal{H}}\right)\,\left(\mathbf{Y}-\left(\boldsymbol{\mathcal{A}}\otimes\mathbf{S}\right) \boldsymbol{\mathcal{H}}\right)^{\rm H}\right\rbrace\nonumber\\
&\,=\mathrm{Tr}\left\lbrace \mathbf{Y}\, \mathbf{Y}^{\rm H}-\mathbf{Y}\, \boldsymbol{\mathcal{H}}^{\rm H}\left(\boldsymbol{\mathcal{A}}^{\rm H}\otimes\mathbf{S}^{\rm H}\right) -\left(\boldsymbol{\mathcal{A}}\otimes\mathbf{S}\right) \boldsymbol{\mathcal{H}}\, \mathbf{Y}^{\rm H}\right.\nonumber\\
&\,\left.\quad\qquad+\left(\boldsymbol{\mathcal{A}}\otimes\mathbf{S}\right) \boldsymbol{\mathcal{H}}\,\boldsymbol{\mathcal{H}}^{\rm H}\left(\boldsymbol{\mathcal{A}}^{\rm H}\otimes\mathbf{S}^{\rm H}\right)\right\rbrace. 
\end{align}
Next, taking derivative of \eqref{eq:tr-obj} with respect to $\mathbf{h}_k,\forall k\in\mathcal{M}$, by
using standard rules for complex-valued matrix differentiation~\cite{complex-matrixbook,golub2012matrix} and setting the resultant to zero, gives below
\begin{align}\label{eq:der-tr-obj-0}
\frac{\partial\mathcal{E}}{\partial\mathbf{h}_k}=&-\mathbf{h}_k^{\rm T}\Big(\mathbf{Y}^{\rm H}\left(\boldsymbol{\mathcal{A}}\otimes\mathbf{S}\right)\left[\mathbf{0}_{N\times\left(k-1\right)N}\,\;\mathbf{I}_N\;\,\mathbf{0}_{N\times\left(M-k\right)N}\right]^{\rm T}\nonumber\\
&\;+\left[\mathbf{0}_{N\times\left(k-1\right)N}\,\;\mathbf{I}_N\;\,\mathbf{0}_{N\times\left(M-k\right)N}\right]\left(\boldsymbol{\mathcal{A}}^{\rm T}\otimes\mathbf{S}^{\rm T}\right)\mathbf{Y}^{*}\Big)\nonumber\\
&\;+\mathbf{h}_k^{\rm T} \bigg(\boldsymbol{\mathcal{H}}^{\rm H}\left(\boldsymbol{\mathcal{A}}^{\rm H}\otimes\mathbf{S}^{\rm H}\right)\left(\boldsymbol{\mathcal{A}}\otimes\mathbf{S}\right)\left[\mathbf{0}_{N\times\left(k-1\right)N}\;\;\mathbf{I}_N\right.\nonumber\\
&\quad\left.\mathbf{0}_{N\times\left(M-k\right)N}\right]^{\rm T} +\left[\mathbf{0}_{N\times\left(k-1\right)N}\,\;\mathbf{I}_N\;\,\mathbf{0}_{N\times\left(M-k\right)N}\right]\nonumber\\
&\;\times\left(\boldsymbol{\mathcal{A}}^{\rm T}\otimes\mathbf{S}^{\rm T}\right)\left(\boldsymbol{\mathcal{A}}^*\otimes\mathbf{S}^*\right)\boldsymbol{\mathcal{H}}^{*}\bigg)=\mathbf{0}_{1\times N}.
\end{align}  

On applying some simplifications to \eqref{eq:der-tr-obj-0}, we obtain the following system of $M$ equations, with $k\in\mathcal{M}$,
\begin{equation}\label{eq:der-tr-obj}
\mathbf{h}_k^{\rm T}\Big(\mathbf{Y}^{\rm H}\,\mathbf{S}_{\rm A}\,\mathbf{X}_k^{\rm T}+\mathbf{X}_k\,\mathbf{S}_{\rm A}^{\rm T}\,\mathbf{Y}^{*}\Big)=2\,\mathbf{h}_k^{\rm T}\,  \boldsymbol{\mathcal{H}}^{\rm H}\,\mathbf{S}_{\rm A}^{\rm H} \,\mathbf{S}_{\rm A}\,\mathbf{X}_k^{\rm T},
\end{equation} 
where $\mathbf{X}_k\triangleq\left[\mathbf{0}_{N\times\left(k-1\right)N}\,\;\mathbf{I}_N\;\,\mathbf{0}_{N\times\left(M-k\right)N}\right],\,\forall\,k\in\mathcal{M},$ is an $N\times NM$ matrix and $\mathbf{S}_{\rm A}\triangleq\left(\boldsymbol{\mathcal{A}}\otimes\mathbf{S}\right)\in\mathbb{C}^{NM\times NM}$. Further, we can write \eqref{eq:der-tr-obj} in a more elegant form given below.
\begin{align}\label{eq:der-tr-obj-id2-0}
\mathbf{h}_k^{\rm H}\,\overline{\mathbf{Y}}_k {=}  \mathbf{h}_k^{\rm H}  \left[\mathbf{0}_{N\times\left(k-1\right)N}\,\;\mathbf{h}_k\mathbf{h}_k^{\rm T}\;\,\mathbf{0}_{N\times\left(M-k\right)N}\right],
\end{align}  
where symmetric matrix $\overline{\mathbf{Y}}_k\in\mathbb{C}^{N\times N},\,\forall k\in\mathcal{M},$ is given by
\begin{eqnarray}\label{eq:Yk}
\overline{\mathbf{Y}}_k\triangleq \frac{\mathbf{Y}^{\rm T}\left(\mathbf{S}_{\rm A}^{\rm H} \,\mathbf{S}_{\rm A}\right)^{-1}\,\mathbf{S}_{\rm A}^*\,\mathbf{X}_k^{\rm H}+\mathbf{X}_k^*\,\mathbf{S}_{\rm A}^{\rm H}\left(\mathbf{S}_{\rm A}^{\rm H}\,\mathbf{S}_{\rm A}\right)^{-1}\mathbf{Y}}{2}.
\end{eqnarray}
It may be noted that for under perfect silent ($\mathrm{off}$) state with $\left|a_0\right|\approx0$, following result holds
\begin{align}\label{eq:der-tr-obj-idb}
\mathbf{S}_{\rm A}^{\rm H} \,\mathbf{S}_{\rm A}=&\,\left(\boldsymbol{\mathcal{A}}\otimes\mathbf{S}\right)^{\rm H}\left(\boldsymbol{\mathcal{A}}\otimes\mathbf{S}\right)=\boldsymbol{\mathcal{A}}^{\rm H} \boldsymbol{\mathcal{A}}\otimes\mathbf{S}^{\rm H} \mathbf{S}\nonumber\\
\approx&\,\abs{a_1}^2\mathbf{I}_M\otimes \frac{p_t\tau_{c_0}}{N}\,\mathbf{I}_N=\frac{\abs{a_1}^2p_t\tau_{c_0}}{N}\,\mathbf{I}_{N M},
\end{align} 
and due to above, $\overline{\mathbf{Y}}_k\approx \frac{\left|a_1\right|^2\,p_t\tau_{c_0}}{2N}\left(\mathbf{Y}^{\rm T}\,\mathbf{S}_{\rm A}^* \mathbf{X}_k^{\rm H}+\mathbf{X}_k^*\,\mathbf{S}_{\rm A}^{\rm H} \,\mathbf{Y}\right)$. Lastly, with $\widetilde{\mathbf{Y}}_k\triangleq \left[\overline{\mathbf{Y}}_k\right]_{1:N,\;\left(k-1\right)N+1\,:\,kN}\in\mathbb{C}^{N\times N}$ defined out of $\overline{\mathbf{Y}}_k$, we obtain the result below,
\begin{align}\label{eq:der-tr-obj-id2} 
\mathbf{h}_k^{\rm H}\,\widetilde{\mathbf{Y}}_k {=} \mathbf{h}_k^{\rm H} \,\mathbf{h}_k\mathbf{h}_k^{\rm T},\quad\forall\, k\in\mathcal{M}.
\end{align}  
As \eqref{eq:der-tr-obj-id2} involves solving a system of $NM$ complex nonlinear equations in $N$ complex entries of $M$ BSC channel vectors ${\mathbf{h}_k},\forall k\in \mathcal{M}$, it can be computationally very expensive if the
antenna array at $\mathcal{R}$ is large ($N\gg 1$) or tags count is high ($M\gg 1$). Therefore, we next present an alternative real-domain representation that can be efficiently solved in parallel, thereby enabling distributed computing, regardless of the size $N$ of the antenna array at $\mathcal{R}$ or tags count $M$. 
\subsubsection{Semi-Closed-Form Expression for LSE}\label{sec:SCF}
Exploiting principal eigenvector approximation, here we present an equivalent transformation for the system of equations defined in \eqref{eq:der-tr-obj-id2} to real domain for obtaining the solution for $\mathcal{O}_{\mathrm L}$, i.e., LSE $\widehat{\mathbf{h}}_k\,\forall k\in\mathcal{M}$. This solution, although not  unique, provides the \textit{globally-minimum} value of objective function $\mathcal{E}$ under phase ambiguity errors. Therefore, the system of $NM$ nonlinear complex equations as defined by \eqref{eq:der-tr-obj-id2} can be  equivalently represented by the $2NM$  \textit{real} equations as defined below by~\eqref{eq:real}.
\begin{align}\label{eq:real}
\mathbf{Z}_k\;\overline{\mathbf{h}}_k=\norm{{\mathbf{h}_k}}^2\,\overline{\mathbf{h}}_k,\quad\forall\, k\in\mathcal{M},
\end{align}
where $\mathbf{Z}_k\triangleq\left[\begin{array}{ccc}
\mathrm{Re}\{\widetilde{\mathbf{Y}}_k\} & \;\,\mathrm{Im}\{\widetilde{\mathbf{Y}}_k\} \\
\mathrm{Im}\{\widetilde{\mathbf{Y}}_k\} &  -\mathrm{Re}\{\widetilde{\mathbf{Y}}_k\} \end{array}    \right]\in\mathbb{R}^{2N\times2N}$ is a real symmetric matrix and $\overline{\mathbf{h}}_k\triangleq\left[\begin{array}{cc}
\mathrm{Re}\{{\mathbf{h}_k}\} \\
\mathrm{Im}\{{\mathbf{h}_k}\} \end{array}\right]$ $\in\mathbb{R}^{2N\times1}$ is a real vector. 
Note that \eqref{eq:real} is obtained after using $\mathbf{h}_k^{\rm T}\,  \mathbf{h}_k^*=\norm{    {\mathbf{h}_k}}^2$ and rewriting the $NM$ complex equations in \eqref{eq:der-tr-obj-id2} for separately finding the underlying real and imaginary terms. As \eqref{eq:real} possesses a conventional eigenvalue problem form, the solution to \eqref{eq:real} in $\overline{\mathbf{h}}_k$ is either given by a zero vector   or by the eigenvector corresponding to the positive eigenvalue $\norm{{\mathbf{h}_k}}^2$ of the matrix $\mathbf{Z}_k$. Thus, we have decoupled solving of this system of $NM$ nonlinear equations to implement $M$ eigenvalue-decompositions in parallel.

It is worth noting here that since $\mathcal{O}_{\rm L}$ involved minimization of $\mathcal{E}=
\norm{\mathbf{Y}-\left(\boldsymbol{\mathcal{A}}\otimes\mathbf{S}\right)\,\boldsymbol{\mathcal{H}}}^2$ and we reduced it to solving $M$ eigenvalue problems, the global minimum value for the objective of $\mathcal{O}_{\rm L}$ is attained at   ${\mathbf{h}_k}=\widehat{\mathbf{h}}_k\triangleq\mathrm{Re}\{\widehat{\mathbf{h}}_k\}+j\,\mathrm{Im}\{\widehat{\mathbf{h}}_k\},\forall\, k\in\mathcal{M}$, where  real and imaginary components for each of the LSE as obtained using the maximum eigenvalue  $\lambda_{\max}\left\lbrace\mathbf{Z}_k\right\rbrace$ of $\mathbf{Z}_k$ are  defined below
\begin{eqnarray}\label{eq:h-GLSE}
\left[\begin{array}{cc}
\mathrm{Re}\{\widehat{\mathbf{h}}_k\} \\
\mathrm{Im}\{\widehat{\mathbf{h}}_k\} \end{array}\right]\triangleq\pm\frac{\sqrt{\lambda_{\max}\left\lbrace\mathbf{Z}_k\right\rbrace}\;\mathbf{v}_{\max}\left\lbrace\mathbf{Z}_k\right\rbrace}{\norm{\mathbf{v}_{\max}\left\lbrace\mathbf{Z}_k\right\rbrace}}\in\mathbb{R}^{2N\times1}.
\end{eqnarray}
Furthermore, since the sign gets  cancelled in the product definition ${\mathbf{h}_k}\,\mathbf{h}_k^{\rm T}$ of $\boldsymbol{\mathcal{H}}$ used in the objective of $\mathcal{O}_{\rm L}$, this LSE $\widehat{\mathbf{h}}_k,\forall k\in\mathcal{M},$ yielding the global minimizer involves an unresolvable phase ambiguity and hence this LSE is  not unique. Without loss of generality, we have considered `$+$' sign for $\widehat{\mathbf{h}}_k,\forall k\in\mathcal{M},$ in \eqref{eq:h-GLSE} from now-onwards in this work.  {Here, it may also be recalled from  \cite{massive-MIMO,TSP19} that theoretically for effective estimation of  the tag $\mathcal{T}_k$ to reader $\mathcal{R}$ channel vector $\mathbf{h}_k$ of size $N$ by an $N$-antenna reader in reciprocal monostatic BSC settings, single pilot count  is sufficient. However, due to the unresolved phase ambiguity as observed from the derived estimate $\widehat{\mathbf{h}}_k$ (cf. \eqref{eq:h-GLSE}, we have considered a larger sized  pilot matrix (i.e., having $N$ orthogonal pilots)   to combat the underlying adverse effects, specifically for high signal-to-noise-ratio (SNR) regimes. Nevertheless, as discoursed in \cite{TSP19}, for low SNR  regime, when the propagation losses are severe during the CE phase, it may be better to allocate all
 $p_t$  to only one of the antennas at $\mathcal{R}$ and try to estimate
the $N$ sized  vector $\mathbf{h}_k$ from the underlying $N$ sized received signal vector, rather than distributing $p_t$ across $N$ antennas of $\mathcal{R}$ for estimating it from an $N \times N$  matrix $\widetilde{\mathbf{Y}}_k$.}
 
\section{Fair Backscattered Throughput Maximization}\label{sec:fair}
In this section we outline the mathematical formulation for the CTM problem, followed by discussion on underlying challenges and the individually-optimal precoder and detector. 

\subsection{Optimization Formulation}
Since, the resource-constrained tags are totally dependent on  energy-rich and computationally-sufficient full-duplex multiantenna reader to carry out their respective BSC, fair allocation of the reader's resources becomes very critical in the multi-tag settings like self-sustainable IoT applications~\cite{New-SP-Mag1}.
To address this timely requirement of next-generation green  wireless networks, the CTM problem for TRX designing under the power constraints at $\mathcal{R}$ can be formulated as
\begin{eqnarray*}\label{eqOPTM1}
	\begin{aligned} 
		\mathcal{O}_{\mathrm{M}}: \;&\underset{\mathbf{f},\,\mathbf{G}}{\text{maximize}}\quad\underset{k\in\mathcal{M}}{\text{minimum}}\quad 
		\mathrm{R}_k\left(\mathbf{f},\mathbf{G}\right),\quad \text{subject to (s.t.)}\\
		&
		({\rm C1}): \lVert\mathbf{f}\rVert^2\le  p_t,\qquad\;\;\;
		({\rm C2}): \lVert\mathbf{g}_k\rVert^2\le 1,\forall k\in\mathcal{M},
	\end{aligned} 
\end{eqnarray*}
where constraint $({\rm C1})$  ensures that the precoder $\mathbf{f}$ is designed satisfying the available transmit (TX) power budget $p_t$ at $\mathcal{R}$. Whereas,  $({\rm C2})$ ensures that all the detectors $\mathbf{g}_k,\forall k\in\mathcal{M},$ are unit-norm receive (RX) beamforming vectors.
\begin{remark}
	\textit{It is worth noting that we have considered a BSC setting with equally-prioritized tags and the goal is to evaluate the minimum guaranteed backscattered throughput, keeping in mind the practical BSC challenges like  CE errors, UAR, and resource-constraints of tags.}
\end{remark}

From \eqref{eqOPTM1}  we observe that $\mathcal{O}_{\mathrm{M}}$ is jointly nonconvex in $\mathbf{f}$ and $\mathbf{G}$. Hence, to reduce the complexity of this TRX designing problem and get some hold on it, we first decouple it into individual precoder and detector optimization problems, which also helps in gaining nontrivial design insights to be later in obtaining the desired joint solution for \eqref{eqOPTM1}. Next investigate optimal precoder for given detector designs and optimal detectors for a known precoder at $\mathcal{R}$.

\subsection{Optimal Receive Beamforming}\label{sec:RX-BF}
We observe that each $\mathrm{R}_k$ is a monotonically increasing function of the underlying SINR $\gamma_k$ for $\mathcal{T}_k$, we note that maximizing  $\underset{k\in\mathcal{M}}{\min}\;\mathrm{R}_k$ is equivalent to maximizing the underlying $\underset{k\in\mathcal{M}}{\min}\;\gamma_k$. Further, for a given precoder design $\mathbf{f}$, the SINR expression $\gamma_k$ can be alternately represented as a generalized Rayleigh quotient~\cite{SOC-SPMag}. Here, since each $\gamma_k$ depends only on its own detector $\mathbf{g}_k$ design, we can optimize them in parallel while satisfying the normalization constraint $({\rm C2})$ at equality. Therefore, for a given precoder $\mathbf{f}$, the optimal detector  design problem can be formulated below as optimizing the RX beamforming separately for each tag $\mathcal{T}_k$, where $k\in\mathcal{M}$,
\begin{eqnarray}\label{eq:OpGM0}
\underset{\mathbf{g}_k:\lVert\mathbf{g}_k\rVert^2=1}{\text{argmax}}\;\;\gamma_k=\frac{\left|\mathbf{g}_{k}^{\rm H}\,\mathbf{h}_{k}\right|^2\,\left|\mathbf{h}_{k}^{\rm T}\mathbf{f}\right|^2}{\sum\limits_{i\in\mathcal{M}_k}\left|\mathbf{g}_{k}^{\rm H}\,\mathbf{h}_{i}\right|^2\,\left|\mathbf{h}_{i}^{\rm T}\mathbf{f}\right|^2+\overline{\sigma}_{\mathrm{w}_{\mathcal{R}}}^2\norm{\mathbf{g}_{k}}^2},
\end{eqnarray}
whose global-optimal solution is ${\mathbf{G}}_{\mathrm{op}}\triangleq\left[{\mathbf{g}}_{\mathrm{op}_1}\;{\mathbf{g}}_{\mathrm{op}_2}\,\ldots\,{\mathbf{g}}_{\mathrm{op}_1}\right]$, whose column entries $\forall k\in\mathcal{M}$ are defined below
\begin{align}\label{eq:opt-RX-0}
\left[\mathbf{G}_{\mathrm{op}}\right]_k\triangleq\frac{\left(\mathbf{I}_N+ \frac{1}{\sigma_{\mathrm{w}_\mathcal{R}}^2}\,\sum\limits_{i=1}^M
\left|\mathbf{h}_{i}^{\rm T}\mathbf{f}\right|^2\mathbf{h}_{i}\,\mathbf{h}_{i}^{\rm H} \right)^{-1}\mathbf{h}_{k}}{\norm{\left(\mathbf{I}_N+ \frac{1}{\sigma_{\mathrm{w}_\mathcal{R}}^2}\,\sum\limits_{i=1}^M
		\left|\mathbf{h}_{i}^{\rm T}\mathbf{f}\right|^2\mathbf{h}_{i}\,\mathbf{h}_{i}^{\rm H} \right)^{-1}\mathbf{h}_{k} }}.
\end{align}
It is worth noting that this detector matrix $\mathbf{G}_{\mathrm{op}}$ is a function of the precoding vector $\mathbf{f}$. Furthermore since we only have the LSE available for  BSC channels $\mathbf{h}_k,\forall k\in\mathcal{M},$ in practical settings the optimal detector $\widehat{\mathbf{G}}_{\mathrm{op}}$ for each tag $\mathcal{T}_k$ as obtained on using \eqref{eq:h-GLSE} in \eqref{eq:opt-RX-0} is given by
\begin{align}\label{eq:opt-RX}
\left[\widehat{\mathbf{G}}_{\mathrm{op}}\right]_k\triangleq\frac{\left(\mathbf{I}_N+ \frac{1}{\sigma_{\mathrm{w}_\mathcal{R}}^2}\,\sum\limits_{i=1}^M
	\left|\widehat{\mathbf{h}}_{i}^{\rm T}\mathbf{f}\right|^2\widehat{\mathbf{h}}_{i}\,\widehat{\mathbf{h}}_{i}^{\rm H} \right)^{-1}\widehat{\mathbf{h}}_{k}}{\norm{\left(\mathbf{I}_N+ \frac{1}{\sigma_{\mathrm{w}_\mathcal{R}}^2}\,\sum\limits_{i=1}^M
		\left|\widehat{\mathbf{h}}_{i}^{\rm T}\mathbf{f}\right|^2\widehat{\mathbf{h}}_{i}\,\widehat{\mathbf{h}}_{i}^{\rm H} \right)^{-1}\widehat{\mathbf{h}}_{k} }}.
\end{align}
 
\subsection{Optimal Transmit Precoding}
Here before focusing on the key features of precoder design for  $\mathcal{R}$  considering multiple tags in CTM formulation, we start with the discussion on optimal precoder for each tag via Lemma~\ref{lem:opt-Tx-Each-tag}. 
\begin{lemma}\label{lem:opt-Tx-Each-tag}
	The optimal precoder for $\mathcal{T}_k$ that maximizes $\mathrm{R}_k$ for a given $\mathbf{G}$ at $\mathcal{R}$ is defined by
	\begin{equation}\label{eq:f-op-k}
	\mathbf{f}_{\rm op}^{(k)}\triangleq\sqrt{p_t}\;\mathbf{v}_{\max}\left\lbrace\left(\boldsymbol{\mathcal{G}}_{\overline{k}}\right)^{-1}{\boldsymbol{\mathcal{G}}}_k\right\rbrace,
	\end{equation}	
	where ${\boldsymbol{\mathcal{G}}}_k\triangleq\mathbf{h}_{k}^*\,\mathbf{h}_{k}^{\rm T}\left|\mathbf{g}_{k}^{\rm H}\,\mathbf{h}_{k}\right|^2$ and  $\boldsymbol{\mathcal{G}}_{\overline{k}}\triangleq\sum\limits_{i\in\mathcal{M}_k}\mathbf{h}_{i}^*\,\mathbf{h}_{i}^{\rm T}\left|\mathbf{g}_{k}^{\rm H}\,\mathbf{h}_{i}\right|^2 +\frac{\overline{\sigma}_{\mathrm{w}_{\mathcal{R}}}^2}{p_t}\norm{\mathbf{g}_{k}}^2\mathbf{I}_N,\forall k\in\mathcal{M}$. 
\end{lemma}

\begin{IEEEproof} We can rewrite the throughput for $\mathcal{T}_k$ defined in \eqref{eq:Rk} as
$\mathrm{R}_k=\left(\frac{\tau-\tau_c}{\tau}\right)\log_2\left(1+\frac{\mathbf{f}^{\rm T}\,\boldsymbol{\mathcal{G}}_k\,\mathbf{f}^*}{\mathbf{f}^{\rm T}\,\boldsymbol{\mathcal{G}}_{\overline{k}}\,\mathbf{f}^*}\right).$ Noting that the logarithm function is monotonic and $\mathrm{R}_k$ being logarithmic transformation of this Rayleigh quotient form,  the optimal precoder for $\mathcal{T}_k$ with known detectors $\mathbf{G}$ at $\mathcal{R}$ is characterized by the generalized eigenvector corresponding to the largest eigenvalue of the matrix set $\left({\boldsymbol{\mathcal{G}}}_k,\boldsymbol{\mathcal{G}}_{\overline{k}}\right)$~\cite{golub2012matrix}. In other words, when $\boldsymbol{\mathcal{G}}_{\overline{k}}$ is invertible, $\mathbf{f}_{\rm op}^{(k)}$ is given by the principal eigenvector corresponding to maximum eigenvalue $\lambda_{\max}\left\lbrace\left(\boldsymbol{\mathcal{G}}_{\overline{k}}\right)^{-1}{\boldsymbol{\mathcal{G}}}_k\right\rbrace$ of the matrix  $\left(\boldsymbol{\mathcal{G}}_{\overline{k}}\right)^{-1}{\boldsymbol{\mathcal{G}}}_k$. 
\end{IEEEproof}

Along with the above result, we next present a key characteristic of the optimal precoder design that leads to the CTM and show how the optimal precoder has to balance among the individually-optimal precoding vectors (cf. \eqref{eq:f-op-k}) and the one that maximizes the fairness.
 
\begin{lemma}\label{lem:equal-R}
	The optimal precoder $\mathbf{f}_{\rm op}$ that maximizes the minimum backscattered rate among the tags is one that balances  between individually optimal precoders for different tags as defined by \eqref{eq:f-op-k} and yields in the equal backscattered rate for all the tags, i.e., $\mathrm{R}_k=\mathrm{R}_1,\forall k\in\mathcal{M}$. 
\end{lemma}
\begin{IEEEproof} 
First let us rewrite $\mathcal{O}_{\mathrm{M}}$ in the form of the following equivalent problem $\mathcal{O}_{\mathrm{M}1}$
	\begin{eqnarray*}\label{eqOPTM2}
		\begin{aligned} 
			\mathcal{O}_{\mathrm{M}1}:  \underset{\mathbf{f},\,\mathbf{G},\,\overline{\mathrm{R}}}{\text{maximize}}\;\overline{\mathrm{R}},\;\,\text{s.t.}\,
			({\rm C1}),({\rm C2}),\;({\rm C3}): \overline{\mathrm{R}}\le\mathrm{R}_k,\forall k\in\mathcal{M}.
		\end{aligned} 
	\end{eqnarray*}
	
	Since the Karush-Kuhn-Tucker (KKT) point~\cite{Baz} characterizes all the candidates for the optimal solution of an optimization problem, next we try to find out the properties of the KKT point for $\mathcal{O}_{\mathrm{M}1}$. In this context, below we define the Lagrangian function $\mathcal{L}_{\mathrm{M1}}$ for problem  $\mathcal{O}_{\mathrm{M}1}$
	\begin{align}\label{eq:Lang-OM1}
	\mathcal{L}_{\mathrm{M1}}&\left(\mathbf{f},\mathbf{G},\overline{\mathrm{R}},\nu_1,\boldsymbol{\nu_2},\boldsymbol{\nu_3}\right)\triangleq\overline{\mathrm{R}}+\nu_{1}\left(p_t-\lVert\mathbf{f}\rVert^2\right)+\nonumber\\
	&\sum\limits_{k\in\mathcal{M}}\left[\nu_{2k}\left(1-\lVert\mathbf{g}_k\rVert^2\right)+\nu_{3k}\left(\mathrm{R}_k-\overline{\mathrm{R}}\right)\right],
	\end{align}
	where $\nu_1,\boldsymbol{\nu_2}\triangleq\left[\nu_{21}\;\;\nu_{22}\;\;\nu_{23}\;\ldots\;\nu_{2M}\right]^{\rm T}\in\mathbb{R}^{M\times 1}_{\ge0},$ and $\boldsymbol{\nu_3}\left[\nu_{31}\;\;\nu_{32}\;\;\nu_{33}\;\ldots\;\nu_{3M}\right]^{\rm T}\in\mathbb{R}^{M\times 1}_{\ge0}$ are respectively the non-negative Lagrange multipliers corresponding to the constraints  $({\rm C1})$,  $({\rm C2})$, and $({\rm C3})$. So, the underlying KKT conditions~\cite{Baz}, except the ones being  defined by the constraints $\mathcal{O}_{\mathrm{M}1}$ and non-negativity of the Lagrange multipliers,  can be written as below
	\begin{subequations}	
		\begin{eqnarray}\label{eq:KKT1-OM1}
		\frac{\partial\mathcal{L}_{\mathrm{M1}}}{\partial\mathbf{f}}=\textstyle\sum_{k\in\mathcal{M}}\;\nu_{3k}\,\frac{\partial\mathrm{R}_k}{\partial\mathbf{f}}-\nu_1\,\mathbf{f}=\mathbf{0}_{N\times1},
		\end{eqnarray}
		\begin{eqnarray}\label{eq:KKT2-OM1}
		\frac{\partial\mathcal{L}_{\mathrm{M1}}}{\partial\mathbf{g}_k}=\nu_{3k}\,\frac{\partial\mathrm{R}_k}{\partial\mathbf{g}_k}-\nu_{2k}\,\mathbf{g}_k=\mathbf{0}_{N\times1},\,\forall k\in\mathcal{M},
		\end{eqnarray}
		\begin{eqnarray}\label{eq:KKT3-OM1}
		\frac{\partial\mathcal{L}_{\mathrm{M1}}}{\partial\overline{\mathrm{R}}}=1-\textstyle\sum_{k\in\mathcal{M}}\;\nu_{3k}=0,
		\end{eqnarray}
		\begin{align}\label{eq:KKT4-OM1}
		\nu_{1}\left(p_t-\lVert\mathbf{f}\rVert^2\right)=&\,\nu_{2k}\left(1-\lVert\mathbf{g}_k\rVert^2\right)\nonumber\\
		=&\,\nu_{3k}\left(\mathrm{R}_k-\overline{\mathrm{R}}\right)=0,\quad\forall k\in\mathcal{M}.
		\end{align}
	\end{subequations}
	
	As noted in Section~\ref{sec:RX-BF}, the optimal detector for each tag is a unit-norm  beamforming vector. Therefore,  $({\rm C2})$ is satisfied at equality for all the tags and hence, the underlying Lagrange multiplier $\nu_{2k}$ has positive shadow price (strictly-positive value at the optimal solution) for all the tags, i.e., $\nu_{2k}>0,\forall k\in\mathcal{M}$. Using this outcome in \eqref{eq:KKT2-OM1}, we notice that for each RX design $\mathbf{g}_k$ to be a non-zero vector, $\nu_{3k}>0$ for all the tags. Using this in \eqref{eq:KKT4-OM1}, we  observe that $({\rm C3})$ is satisfied at equality with  $\mathrm{R}_k=\overline{\mathrm{R}} \;(\text{or }\mathrm{R}_1),\forall k\in\mathcal{M}$, which completes the proof.
\end{IEEEproof}

Using these properties of the individually optimal precoder and detector designs as outlined in this section, next we present an efficient low-complexity iterative algorithm to yield the jointly-optimal TRX design for the reader that leads to the CTM among the equally-prioritized tags.


\section{Proposed Solution Methodology}\label{sec:sol}
Recalling the developments from previous section, we can summarize that the optimal TRX design for CTM can be equivalently represented as the precoder designing problem for $\mathcal{R}$ with the underlying optimal detector matrix $\widehat{\mathbf{G}}_{\mathrm{op}}$ being defined as a function of this precoder and LSE given by \eqref{eq:opt-RX}. However, the challenging part is that even this reduced precoder designing problem is nonconvex and in fact it involves computationally-expensive operations like matrix inversions (due to the detector definition used) and complex differentiations. To efficiently handle these challenges, here we outline a novel solution methodology that  not only provides fast convergence to the near-optimal solution for the CTM problem $\mathcal{O}_{\rm M}$, but has very low-complexity which suits the requirements of low-power (resource-constrained) applications like BSC and IoT. In particular, we present the key details for this proposed algorithm, followed by its step-by-step implementation and complexity analysis discussion. But before that we next discuss the asymptotic properties of the optimal TRX design that will be used later in Section~\ref{sec:low} as key input parameters for the proposed iterative algorithm in achieving fast convergence.

\subsection{Asymptotically-Optimal TRX designs}\label{sec:asymp-opt}
We present the asymptotically-optimal TRX designs for both low and high SNR regimes. These designs will be then later used as inputs to the proposed algorithm. 

\subsubsection{Optimal TRX for Low SNR Applications}\label{sec:lowSNR}
This scenario is more common in BSC settings that are significantly affected by the doubly-near-far problem as faced by the tags placed at larger distances. In other words, due to the involvement of the product channels in BSC settings, the wireless  propagations losses are more prominent than the conventional systems and as a result the end-to-end SNR are relatively low. However, this is practically not a major concern because the BSC are generally meant for low-power and lower bit-rate applications like sustainable IoT with EH nodes.   Therefore, under low-SNR regime, we notice that the noise power is relatively stronger than the interference. In other words, the following approximation holds $\forall k\in\mathcal{M}$, 
\begin{equation}\label{eq:L-I}
\sum\limits_{i\in\mathcal{M}_k}\left|\mathbf{g}_{k}^{\rm H}\,\mathbf{h}_{i}\right|^2\,\left|\mathbf{h}_{i}^{\rm T}\mathbf{f}\right|^2+\overline{\sigma}_{\mathrm{w}_{\mathcal{R}}}^2\norm{\mathbf{g}_{k}}^2\approx{\overline{\sigma}_{\mathrm{w}_{\mathcal{R}}}^2}\norm{\mathbf{g}_{k}}^2.
\end{equation} 

Hence, as in this case the interference at any tag $\mathcal{T}_k$ of interest due to the backscattered signals from all the other tags $\mathcal{T}_i,\forall i\in\mathcal{M}_k,$ is relatively very low in comparison to the received AWGN and therefore can be ignored in comparison to the noise power. Therefore, the optimal RX beamforming design reduces to the maximum ratio combining (MRC), i.e., $\mathbf{g}_{k}=\frac{\widehat{\mathbf{h}}_k}{\norm{\widehat{\mathbf{h}}_k}},\forall k\in\mathcal{M},$ yielding the below mentioned SNR approximation $\gamma_{\mathrm{m}_k}$ for $\gamma_k$ to support low SNR applications, 
\begin{align}
\gamma_k=&\,\frac{\left|\mathbf{g}_{k}^{\rm H}\,\mathbf{h}_{k}\,\mathbf{h}_{k}^{\rm T}\,\mathbf{f}\right|^2}{\sum\limits_{i\in\mathcal{M}_k}\left|\mathbf{g}_{k}^{\rm H}\,\mathbf{h}_{i}\,\mathbf{h}_{i}^{\rm T}\,\mathbf{f}\right|^2+\overline{\sigma}_{\mathrm{w}_{\mathcal{R}}}^2\norm{\mathbf{g}_{k}}^2}\nonumber\\
\approx&\,\frac{\left|\mathbf{g}_{k}^{\rm H}\,\mathbf{h}_{k}\right|^2\,\left|\mathbf{h}_{k}^{\rm T}\mathbf{f}\right|^2}{ {\overline{\sigma}_{\mathrm{w}_{\mathcal{R}}}^2}\norm{\mathbf{g}_{k}}^2}\,\stackrel{(r1)}{\le}\, \gamma_{\mathrm{m}_k}\left|\widehat{\mathbf{h}}_{k}^{\rm T}\mathbf{f}\right|^2,\forall\,k\in\mathcal{M},
\end{align} 
where ${\gamma}_{\mathrm{m}_k}\triangleq\frac{\norm{\widehat{\mathbf{h}}_{k}}^2}{\overline{\sigma}_{\mathrm{w}_{\mathcal{R}}}^2}$ and $(r1)$ is obtained noting that we only have the information available on LSE $\widehat{\mathbf{h}}_k$, and not on the actual BSC channel vector $\mathbf{h}_k$.
Following above, we note that the optimal precoder design has to maximize the underlying minimum scaled received power $\underset{k\in\mathcal{M}}{\min}\;\gamma_{\mathrm{m}_k}\left|\widehat{\mathbf{h}}_{k}^{\rm T}\mathbf{f}\right|^2$. Here on using the matrix definition $\boldsymbol{\mathcal{F}}\triangleq\mathbf{f}\,\mathbf{f}^{\rm H}$ and ignoring the rank-one constraint on $\boldsymbol{\mathcal{F}}$, the equivalent SDR for  optimizing this variable  $\boldsymbol{\mathcal{F}}$ can be formulated below as $\mathcal{O}_{\mathrm{S_L}}$.
\begin{eqnarray*}\label{eq-EFM-L}
	\begin{aligned} 
	\mathcal{O}_{\mathrm{S_L}}:\;& \underset{\boldsymbol{\mathcal{F}}\succeq0,\;\mathcal{P}}{\text{maximize}} \;  \mathcal{P},\;\;\text{s t.}\;
		({\rm C4}): \mathrm{Tr}\left\lbrace\boldsymbol{\mathcal{F}}\right\rbrace\le  p_t,\\
		&
		({\rm C5}): \mathrm{Re}\left\lbrace\widehat{\mathbf{h}}_{k}^{\rm T}\,\boldsymbol{\mathcal{F}}\,\widehat{\mathbf{h}}_{k}^*\right\rbrace \ge\frac{\mathcal{P}\, {\gamma}_{\mathrm{m}_1}}{{\gamma}_{\mathrm{m}_k}},\forall k\in\mathcal{M}. 
	\end{aligned} 
\end{eqnarray*} 
We notice that  $\mathcal{O}_{\mathrm{S_L}}$  is a convex problem in the optimization variable $\boldsymbol{\mathcal{F}}$ that  can be solved using the CVX toolbox~\cite{cvx} in \texttt{Matlab}. Next to ensure that this precoding solution, as denoted by $\boldsymbol{\mathcal{F}}_{\rm L}$, satisfies the rank-one constraint, we deploy the conventional randomization process~\cite{Multicast-Rand} and eventually obtain the optimal TX precoder $\mathbf{f}_{\mathrm L}$ for low SNR settings. Lastly, using this precoder, the optimal detector matrix $ \mathbf{G}_{\rm L}$ is finally obtained by substituting $\mathbf{f}=\mathbf{f}_{\mathrm L}$ in \eqref{eq:opt-RX}.
 

\subsubsection{TRX  Design Under High-SNR Regime}\label{sec:high}
We recall that for high SNR regimes~\cite[eq. (14)]{SOC-SPMag}, the optimal minimum mean square error (MMSE) based RX beamforming defined  in \eqref{eq:opt-RX} reduces to the below zero-forcing (ZF) based design 
\begin{eqnarray}
\mathbf{g}_{\mathrm{H}_k}=\frac{\left[\mathbf{G}_{\rm Z}\right]_k}{\norm{\left[\mathbf{G}_{\rm Z}\right]_k}},\forall k\in\mathcal{M},\;\text{with }\;\mathbf{G}_{\rm Z}=\widehat{\mathbf{H}}\,\left(\widehat{\mathbf{H}}^{\rm H}\widehat{\mathbf{H}}\right)^{-1},
\end{eqnarray}  
where $\widehat{\mathbf{H}}\triangleq\left[\widehat{\mathbf{h}}_1\;\;\widehat{\mathbf{h}}_2\;\;\widehat{\mathbf{h}}_3\;\ldots\;\widehat{\mathbf{h}}_M\right]\in\mathbb{C}^{N\times M}.$ 
Thus, following the discussion in Section~\ref{sec:lowSNR}, we note that with  ${\gamma}_{\mathrm{z}_k}\triangleq\frac{1}{\overline{\sigma}_{\mathrm{w}_{\mathcal{R}}}^2{\norm{\big[\mathbf{G}_{\rm Z}\big]_k}}^2},\forall\,k\in\mathcal{M}$, the CTM problem  reduces to maximizing the minimum scaled backscattered SNR $\gamma_{\mathrm{z}_k}\left|\widehat{\mathbf{h}}_{k}^{\rm T}\mathbf{f}\right|^2$ among the tags by optimally designing the precoder $\mathbf{f}$ at $\mathcal{R}$. Now similar to as in case of low SNR regime, here again we can deploy SDR followed by randomization and MMSE filtering operations, but with ${\gamma}_{\mathrm{m}_k}$ substituted for ${\gamma}_{\mathrm{z}_k}$ to eventually obtain the optimal TRX design $\left(\mathbf{f}_{\rm H},\mathbf{G}_{\rm H}\right)$ for the high SNR case. Below we summarize the key take-away message on asymptotically-optimal TRX designs.
 
\begin{remark}\label{rem:high}
	\textit{For the asymptotic (either low or high SNR) scenarios, the TRX optimization reduces to first obtaining the precoder that maximizes the minimum scaled backscattered power maximization from the tags  (which is analogous to TX beamforming for optimal  multicasting). Then this precoding vector is used to obtain MMSE-based RX beamforming design.}
\end{remark}


\begin{algorithm}[!t]
	{\small \caption{Pseudocode for finding asymptotically-optimal TRX design $\left(\mathbf{f}_{\rm A},\mathbf{G}_{\rm A}\right),\forall\rm{A}\in\left\lbrace L,H\right\rbrace$.}\label{Algo:Asy-Opt}
		\begin{algorithmic}[1]
			\Require LSE $\widehat{\mathbf{h}}_k,\forall k\in\mathcal{M},$ known system parameters like $p_t,M,N,\tau,\tau_c,\overline{\sigma}_{\mathrm{w}_{\mathcal{R}}}^2$, choice $\rm{ch}\in\{1,2\}$, sample count $K$.
			\Ensure Optimal TRX design for low or high SNR cases, i.e.,  $\left(\mathbf{f}_{\rm L},\mathbf{G}_{\rm L}\right)$ for  $\rm{ch=1}$ or $\left(\mathbf{f}_{\rm H},\mathbf{G}_{\rm H}\right)$ for $\rm{ch=2}$.
			\If{$\left({\rm ch}=1\right)$}\Comment{Initialising low SNR case}
			\State Set $\widehat{\gamma}_k={\gamma}_{\mathrm{m}_k},\forall\,k\in\mathcal{M}.$
			\ElsIf{$\left({\rm ch}=2\right)$}\Comment{Initialising high SNR case}
			\State Set $\widehat{\gamma}_k={\gamma}_{\mathrm{z}_k},\forall\,k\in\mathcal{M}.$
			\EndIf
			\State Solve the convex problem $\mathcal{O}_{\mathrm{S}}$ using CVX in \texttt{Matlab} and set its global-optimal solution to $\boldsymbol{\mathcal{F}}_{\rm op}$ \label{stp:SDR}\Comment{SDR}
			\begin{eqnarray*}\label{eq-SDR}
				\begin{aligned} 
					\mathcal{O}_{\mathrm{S}}:\;&\,\underset{\boldsymbol{\mathcal{F}}\succeq0,\;\mathcal{P}}{\text{maximize}}\;\mathcal{P},\quad\,\text{s. t.}\;\;({\rm C4}): \mathrm{Tr}\left\lbrace\boldsymbol{\mathcal{F}}\right\rbrace\le  p_t,\\
					&\,					({\rm C6}): \mathrm{Re}\left\lbrace\widehat{\mathbf{h}}_{k}^{\rm T}\,\boldsymbol{\mathcal{F}}\,\widehat{\mathbf{h}}_{k}^*\right\rbrace \ge\frac{\mathcal{P}\, \widehat{\gamma}_{1}}{\widehat{\gamma}_k},\,\forall k\in\mathcal{M},\qquad
				\end{aligned} 
			\end{eqnarray*}
			\State Apply eigenvalue  decomposition to obtain: $\boldsymbol{\mathcal{F}}_{\rm op}=\mathbf{U}\boldsymbol{\Lambda}\mathbf{U}^{\rm H}$, and set $m=1$.\Comment{Randomization process starts}\label{step:RandPs}
			\State Generate $K$ independent unit-variance zero-mean circularly-symmetric complex Gaussian vectors $\mathbf{x_a}_k$  and  $K$ independent uniformly distributed random vectors $\boldsymbol{\theta}_k$   on $\left[0,2\pi\right)$, where both these $K$ vectors are $N\times1$ in size. 
			\State Set $\left[\mathbf{x_b}_k\right]_i=\mathrm{e}^{\left[\boldsymbol{\theta}_k\right]_i},$ where $i=1,2,\ldots,N$ and $k=1,2,\ldots,K.$ 
			\Do 	
			\State Set $\mathbf{f}_{1,m}=\sqrt{p_t}\,\frac{\mathbf{U}\boldsymbol{\Lambda}^{\frac{1}{2}}\mathbf{x_a}_m}{\norm{\mathbf{U}\boldsymbol{\Lambda}^{\frac{1}{2}}\mathbf{x_a}_m}},\;\mathbf{f}_{2,m}=\sqrt{p_t}\,\frac{\mathrm{diag}\left\lbrace\boldsymbol{\mathcal{F}}_{\rm op}\right\rbrace\odot\mathbf{x_b}_m}{\norm{\mathrm{diag}\left\lbrace\boldsymbol{\mathcal{F}}_{\rm op}\right\rbrace\odot\mathbf{x_b}_m}},\;\text{ and }\;\mathbf{f}_{3,m}=\sqrt{p_t}\,\frac{\mathbf{U}\boldsymbol{\Lambda}^{\frac{1}{2}}\mathbf{x_b}_m}{\norm{\mathbf{U}\boldsymbol{\Lambda}^{\frac{1}{2}}\mathbf{x_b}_m}}$.	 
			\State \parbox[t]{\dimexpr\linewidth-\algorithmicindent-\algorithmicindent\relax}{Set $\mathrm{R}_{\mathrm{M}}^{\left(n,m\right)}=\underset{k\in\mathcal{M}}{\min}\quad {\widehat{\gamma}_k} \;\mathrm{Re}\left\lbrace\widehat{\mathbf{h}}_{k}^{\rm T}\,\mathbf{f}_{n,m}\,\mathbf{f}_{n,m}^{\rm H}\,\widehat{\mathbf{h}}_{k}^*\right\rbrace,\quad\forall n=\{1,2,3\}$.} 
			\State Update $m=m+1$.
			\doWhile{$m\le K$}\Comment{Randomization process ends} 
			\State Set $\left(n_{\rm op},m_{\rm op}\right)\triangleq\underset{n=\{1,2,3\},\,m=\{1,2,\ldots,K\}}{\mathrm{argmax}}\quad\mathrm{R}_{\mathrm{M}}^{\left(n,m\right)}$.\label{step:RandPe}				
			\If{$\left({\rm ch}=1\right)$}\Comment{Output for low SNR case}
			\State Set $\mathbf{f}_{\rm L}=\mathbf{f}_{n_{\rm op},m_{\rm op}}\text{ and obtain }\mathbf{G}_{\rm L}$ by substituting $\mathbf{f}_{\rm L}$ in~\eqref{eq:opt-RX}.
			\ElsIf{$\left({\rm ch}=2\right)$}\Comment{Output for high SNR case}
			\State Set $\mathbf{f}_{\rm H}=\mathbf{f}_{n_{\rm op},m_{\rm op}}\text{ and obtain }\mathbf{G}_{\rm H}$ by substituting $\mathbf{f}_{\rm H}$ in~\eqref{eq:opt-RX}.
			\EndIf
		\end{algorithmic}
	}
\end{algorithm} 

\subsubsection{Algorithmic Implementation}  
Noting the similarity in the solution methodology  for obtaining the low and high SNR based approximation for the optimal TRX design, we online the step-by-step procedure for obtaining the two asymptotically-optimal design via Algorithm~\ref{Algo:Asy-Opt}. It starts with solving the  convex SDR-based  optimization problem  $\mathcal{O}_{\mathrm{S}}$ as outlined in step~\ref{stp:SDR} with appropriately defined input  parameters via steps 1 to 4. Thereafter, the underlying precoding solution is made to satisfy the rank-one constraint via the randomization process~\cite{Multicast-Rand} as given by steps \ref{step:RandPs} to \ref{step:RandPe} of Algorithm~\ref{Algo:Asy-Opt}, which finally returns the optimal TX precoder, i.e.,  $\mathbf{f}_{\rm L}$ or $\mathbf{f}_{\rm H}$. The complexity of the randomization process depends on the number of random samples  $K$ used for generating the potential candidates for the optimal precoder and  it needs to be selected carefully based on the desired tradeoff between the solution-quality and computational-complexity.
 
\subsection{Novel Iterative Algorithm for Jointly-Optimal TRX Design}\label{sec:low}
Before presenting the details involved in the implementation of proposed iterative algorithm along with its potential merits,  we first discourse the basic ideology behind its adoption.

\subsubsection{Basic Ideology}
We use the Nelder--Mead (NM) method~\cite{Baz} based low-complexity iterative algorithm that does not require the explicit computation of complex derivatives for sum-backscattered-throughput with respect to the optimization variables. There are four main concepts that have been used.  First, we exploit the knowledge on  MMSE-based  RX beamforming design from Section~\ref{sec:RX-BF} to reduce the TRX designing problem to the precoder-designing one.  

Next since, the performance of NM method is strongly influenced by the starting point, we choose it as the better one (in terms of higher common-backscattered-throughput) between  $\mathbf{f}_{\rm L}$ and $\mathbf{f}_{\rm H}$, i.e.,  the two asymptotically-optimal precoder designs as derived in Section~\ref{sec:asymp-opt} for it. 

Third concept is that in each iteration of the NM method, the next starting point (precoder) is selected based on improving the rate of the bottleneck tag having the lowest backscattered rate in the current iteration. So,  we try to appropriately move the   precoder obtained from the previous iteration in the maximum ratio transmission (MRT) direction for weakest tag.

Last concept is that with optimal detector satisfying the MMSE design, the precoder should satisfy the Lemma~\ref{lem:equal-R} result stating that at the optimal (or termination of the algorithm) each tag should have same backscattered rate. So, in each iteration we keep track of  the standard deviation $\sigma_{\mathrm{R}}$ among the backscattered-throughputs of tags and check if it is below an  acceptable tolerance $\epsilon$. Henceforth, when the improvement is negligible (i.e., $\sigma_{\mathrm{R}}$ is below $\epsilon$) or when the sufficient number of iterations  $\rm{it}_{\max}$ are completed, the algorithm  terminates with the proposed TRX design $\left(\mathbf{f}_{\rm J},\mathbf{G}_{\rm J}\right)$  yielding significantly higher max-min rate than the exiting designs.   

\begin{algorithm}[!t]
	{\small \caption{Iterative algorithm for the jointly-optimal TRX design $\left(\mathbf{f}_{\rm J},\mathbf{G}_{\rm J}\right)$  for CTM in $\mathcal{O}_{\rm M}$.}\label{Algo:J-Opt}
		\begin{algorithmic}[1]
			\Require
			\parbox[t]{\dimexpr\linewidth-\algorithmicindent-\algorithmicindent-\algorithmicindent\relax}{LSE $\widehat{\mathbf{h}}_k,\forall k\in\mathcal{M},$ for all the BSC channel vectors, initial precoder $\mathbf{f}^{(0)}$, known system parameters like $p_t,M,N,\tau,\tau_c,\overline{\sigma}_{\mathrm{w}_{\mathcal{R}}}^2$, along with  maximum iteration count $\rm{it}_{\max}$, and acceptable tolerance $\epsilon$.\strut}\medskip 
			\Ensure Jointly-optimal  TRX design $\left(\mathbf{f}_{\rm J},\mathbf{G}_{\rm J}\right)$ yielding max-min rate $\mathrm{R}_{\rm J}$ for acceptable iterations and tolerance.
			\State Set $\mathrm{it}=1,\;\mathrm{R}_{\mathrm{J}}=0,\;$ and $\;\mathbf{f}^{(\rm it)}=\mathbf{f}^{(\rm 0)}$.\Comment{Initialization -- Outer Loop}
			\Do\Comment{Iteration -- Outer Loop}			
			\State 
			\parbox[t]{\dimexpr\linewidth-\algorithmicindent-\algorithmicindent-\algorithmicindent\relax}{Apply NM method with $\mathbf{f}^{(\rm it)}$ as  starting point for minimizing the standard deviation $\sigma_{\mathrm{R}}$ among  $\mathrm{R_k}\left(\mathbf{f},\widehat{\mathbf{G}}_{\mathrm{op}}\right),\forall k\in\mathcal{M}$, as obtained on substituting~\eqref{eq:opt-RX} in~\eqref{eq:Rk}, by  optimizing precoding vector  $\mathbf{f}$.\strut}\medskip 
			\State \parbox[t]{\dimexpr\linewidth-\algorithmicindent\relax}{Assign the resulting output rates of above NM method to $\mathrm{R}_{k}^{\rm(it)}$ along with  corresponding $\sigma_{\mathrm{R}}$ to  $\sigma_{\mathrm{R}}^{\rm(it)}$.}
			\State Set $k_0\triangleq\underset{k\in\mathcal{M}}{\mathrm{argmin}}\;\;\mathrm{R}_{k}^{\rm(it)}$.	 
			\If{$\left(\mathrm{R}_{k_0}^{\rm(it)}\ge\mathrm{R}_{\mathrm{J}}\right)$}
			\State \parbox[t]{\dimexpr\linewidth-\algorithmicindent\relax}{Assign optimal precoder from current run of NM method to $\mathbf{f}_{\rm J}$  along with  corresponding $\sigma_{\mathrm{R}}^{\rm(it)}$ to  $\sigma_{\mathrm{R_J}}$.}
			\State Set  $\mathrm{R}_{\mathrm{J}}=\mathrm{R}_{k_0}^{\rm(it)}$ and obtain $\mathbf{G}_{\rm J}$ by using $\mathbf{f}=\mathbf{f}_{\rm J}$ in~\eqref{eq:opt-RX}. 
			\State Set $n$ to uniformly-distributed integer between $1$ and $N$.
			\State
			\parbox[t]{\dimexpr\linewidth-\algorithmicindent-\algorithmicindent-\algorithmicindent\relax}{ Generate a vector of $n$ values sampled uniformly at random, without replacement, ranging  from integers between $1$ to $N$, and store them in an array $\boldsymbol{\varphi}$. Therefore, $\boldsymbol{\varphi}\in\mathbb{Z}^{n\times1}$.\strut}\medskip 
			\State Set $m=1,\;$ $\boldsymbol{\alpha}=\mathbf{0}_{N\times1}$.\Comment{Initialization -- Inner Loop}
			\Do 	\Comment{Iteration -- Inner Loop}
			\State
			\parbox[t]{\dimexpr\linewidth-\algorithmicindent-\algorithmicindent-\algorithmicindent\relax}{Generate a uniformly distributed number  between $0$ and $1$, and assign it to  $\left[\boldsymbol{\varphi}\right]_m$th entry of the weight array $\boldsymbol{\alpha}$ $\left(\text{in other words, assign it to } \left[\boldsymbol{\alpha}\right]_{\left[\boldsymbol{\varphi}\right]_m}\right)$.\strut}\medskip 
			\State Update $m=m+1$.
			\doWhile{$m\le n$} \Comment{Termination -- Inner Loop}
			\State Update $\mathrm{it}=\mathrm{it}+1$
			\State Update $\mathbf{f}^{(\rm it)}=\sqrt{p_t}\, \frac{\boldsymbol{\alpha}\odot\widehat{\mathbf{h}}_{k_0}^*+\left(\mathbf{1}_{N\times1}-\boldsymbol{\alpha}\right)\odot\mathbf{f}^{(\rm it-1)}}{\norm{\boldsymbol{\alpha}\odot\widehat{\mathbf{h}}_{k_0}^*+\left(\mathbf{1}_{N\times1}-\boldsymbol{\alpha}\right)\odot\mathbf{f}^{(\rm it-1)}}}$.  
			\EndIf			
			\doWhile{$\left(\left[\sigma_{\mathrm{R}}^{\rm(it)}\ge\epsilon\right]\wedge\left[\rm{it}\le\rm{it}_{\max}\right]\right)$}\Comment{Termination -- Outer Loop}
		\end{algorithmic}
	}
\end{algorithm}  
\subsubsection{Implementation Details}
The detailed steps involved in the implementation of the proposed iterative solution methodology are outlined in Algorithm~\ref{Algo:J-Opt}. It requires the LSE for BSC channels  given by \eqref{eq:h-GLSE} along with the basic known system parameters as its inputs, to eventually yield the near-optimal TRX design with an acceptable tolerance $\epsilon$. Each iteration of Algorithm~\ref{Algo:J-Opt} involves the computation of optimal precoding vector $\mathbf{f}$ using the NM method with the underlying rate definition in~\eqref{eq:Rk} using the detector matrix $\mathbf{G}=\widehat{\mathbf{G}}_{\mathrm{op}}$ as defined  in~\eqref{eq:opt-RX}. As NM method requires a starting point, we  use ${\mathbf{f}}^{(0)}$ as the initial precoder and then try to minimize the standard deviation  among the backscattered rates of the tags by optimizing the precoder $\mathbf{f}$ at $\mathcal{R}$ while setting $\mathbf{G}=\widehat{\mathbf{G}}_{\mathrm{op}}$. Accordingly, we select the starting precoder ${\mathbf{f}}^{(0)}$ as ${\mathbf{f}}_{\rm L}$ if $\underset{k\in\mathcal{M}}{\min}\;\mathrm{R}_k\left(\mathbf{f}_{\rm L},\mathbf{G}_{\rm L}\right)\ge\underset{k\in\mathcal{M}}{\min}\;\mathrm{R}_k\left(\mathbf{f}_{\rm H},\mathbf{G}_{\rm H}\right)$. Otherwise, we use ${\mathbf{f}}^{(0)}={\mathbf{f}}_{\rm H}$. This NM method is iteratively called for at-most $\rm{it}_{\max}$ times, with different precoding vector $\mathbf{f}^{(\rm it)}$ as the starting point for each iteration `${\rm it}$'. The  precoder $\mathbf{f}^{(\rm it)}$ for iteration `${\rm it}$', as designed smartly to aid the bottleneck tag $\mathcal{T}_{k_0}$, is given by
\begin{equation}
\mathbf{f}^{(\rm it)}=\sqrt{p_t}\, \frac{\boldsymbol{\alpha}\odot\widehat{\mathbf{h}}_{k_0}^*+\left(\mathbf{1}_{N\times1}-\boldsymbol{\alpha}\right)\odot\mathbf{f}^{(\rm it-1)}}{\norm{\boldsymbol{\alpha}\odot\widehat{\mathbf{h}}_{k_0}^*+\left(\mathbf{1}_{N\times1}-\boldsymbol{\alpha}\right)\odot\mathbf{f}^{(\rm it-1)}}},
\end{equation}
where $k_0$ is the index of the tag with lowest backscattered rate in the iteration `${\rm it}$' and $\boldsymbol{\alpha}$ is the weighting vector that relates the relative importance of the previous precoding vector and the MRT direction dedicated to the weakest tag in the current iteration. Note that Algorithm~\ref{Algo:J-Opt} returns a suboptimal TRX design yielding much higher minimum or common backscattered-throughput among the tags  than both of the two asymptotically-optimal joint designs. Here, the number of NM-method restarts or iterations $\rm{it}_{\max}$ and tolerance $\epsilon$ need to be judiciously selected based on the desired tradeoff between computational complexity and the solution $\left(\mathbf{f}_{\rm J},\mathbf{G}_{\rm J}\right)$ quality. 
\begin{remark}
\textit{The proposed optimal precoder $\mathbf{f}_{\rm J}$ is obtained via the derivative-free NM method by iteratively updating the better between  two asymptotically-optimal ones in the direction of the channel vector corresponding to the underlying weakest tag in terms of common-backscattered-throughput, while setting the detector $\mathbf{G}_{\rm J}$ based on the MMSE design.}
\end{remark}

\subsubsection{Key Merits} This proposed iterative algorithm is computationally-efficient as it does not involve any complicated operations like complex differentiation or solving of nonlinear equation with matrix inversions.  This is important, especially for low-power networking scenarios like BSC and IoT, because solving large system of nonlinear equations for massive array sized reader with large tags count  can be computationally very  expensive along with the calculation of   gradients due to the involvement of matrix-inverse operations in the MMSE-based detector definition. The fundamental advantage of this derivative-free NM method is that provided a good initial starting point in the form of asymptotically-optimal precoder $\mathbf{f}^{(0)}$, in  most scenarios it succeeds in obtaining an acceptable sub-optimal solution in a relatively small number of function evaluations as compared to most meta-heuristics  in the literature. Furthermore, as  verified in Section~\ref{sec:verify-conv} via extensive simulations, Algorithm~\ref{Algo:J-Opt} provides fast convergence to the optimal TRX design under an acceptable tolerance $\epsilon$ or yields a near-optimal design after reasonable $\mathrm{it}_{\max}$ iterations. 
\begin{figure}[!t]	
		\centering\includegraphics[width=3.48in]{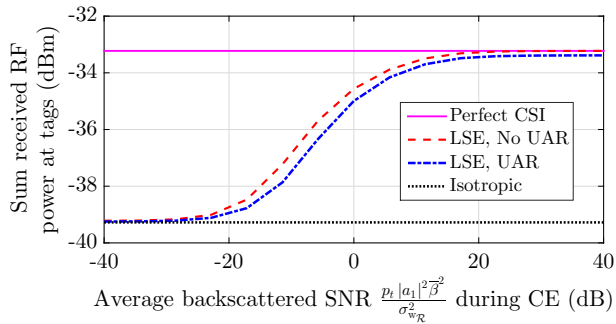}		
		\caption{Validating the quality of proposed LSE in terms of sum received power for different SNRs against perfect CSI-based and isotropic transmissions.}
		\label{fig:CE}  
\end{figure}     
\section{Numerical Results}\label{sec:res}   
In this section we conduct a detailed numerical investigation to quantify the efficacy of our proposed CE protocol and optimal TRX designs. Unless explicitly stated, the default system parameter values are $N=4, M=4$, $p_t=30$dBm, $\sigma_{\mathrm{w}_{\mathcal{R}}}^2=-140$dBm, $K=10\,NM, \mathrm{it}_{\max}=15,\epsilon=10^{-3},\tau=10^{-3}$s$,\tau_{c_k}=\frac{\tau}{10(M+1)}$s and $\beta_k=\left(\frac{3\times 10^8}{4\pi f}\right)^2d_k^{-\varrho},\forall k\in\mathcal{M}$, where $f=915$MHz  is the transmit frequency, $d_k$ is $\mathcal{R}$-to-$\mathcal{T}_k$ distance, and $\varrho=3$ is  path loss exponent. {Noting practical settings for the refection coefficients in BSC, we use $a_0=0.1,a_1=0.78$, and $\overline{a}=0.3162$~\cite{Bistatic-BCS,TSP19}.} {Regarding implementation of UAR, we consider that the entries of $\mathbf{H}_{\rm U}$ are zero-mean circularly
symmetric complex Gaussian random variables with variance $\sigma_{\rm H_U}^2=-90$dBm.} {Due to the consideration of full-duplex multiantenna reader~\cite{TSP19,Impinj} and semi-passive tags~\cite{semi-passive}, in these simulations, we assume that  $M$ tags are uniformly deployed over a square field of length $L=100$m with $\mathcal{R}$ being placed at its center.} Lastly, all the CTM performance results here have been generated after taking average over $10^3$ independent channel realizations. 
\begin{figure}[!t] 
		\centering\includegraphics[width=3.48in]{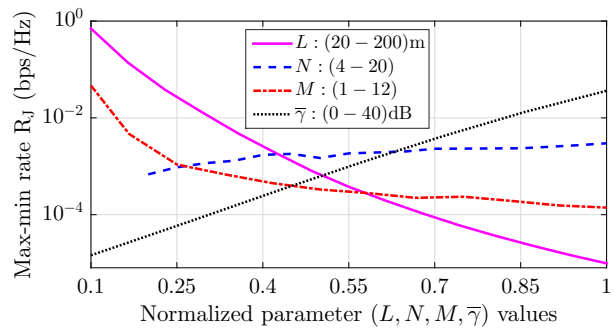}
		\caption{ Variation of max-min rate $\mathrm{R_J}$ as returned by Algorithm~\ref{Algo:J-Opt} for different  parameters like field size $L$, array size $N$, tags count $M$, average SNR $\overline{\gamma}$.}
		\label{fig:Norm_R}   
\end{figure} 
\subsection{Validation of the Proposed LSE for the BSC Channels} 
With the help of Fig.~\ref{fig:CE} we verify the performance of proposed LSE $\widehat{\mathbf{h}}_k,\forall k\in\mathcal{M},$ as defined by~\eqref{eq:h-GLSE}, against increasing values for average backscattered SNR $\frac{p_t \left|a_1\right|^2\overline{\beta}^2}{\sigma_{\mathrm{w}_{\mathcal{R}}}^2}$, where $\overline{\beta}\triangleq\frac{1}{M}\sum_{k=1}^{M}\beta_k$. Here, we have considered the average sum-received-power among the tags  as performance validation metric for estimating the goodness of  proposed LSE, for both with and without UAR. Other than the sum-received power at the tags using LSE with and without UAR, we have also plotted the perfect CSI-based and isotropic (requiring no CE) transmission  cases in Fig.~\ref{fig:CE}  to respectively represent the upper and lower bounds on this performance metric.

From Fig.~\ref{fig:CE}, we notice that the quality of  proposed LSE  $\widehat{\mathbf{h}}_k,\forall k\in\mathcal{M},$ improves with increasing SNR because the underlying CE errors gets reduced for both with and without UAR. In fact for SNRs larger than $20$dB,  the sum-received-power at the tags as obtained using LSE for the scenarios with no UAR is approximately same as the one achieved with perfect CSI availability. {However, for the practical scenarios suffering from UAR, this sum-received-power is $0.15$dBm lower than the maximum achievable even for very high SNRs. In contrast, for low SNR regime, the performance of LSE, with or without UAR, is similar to that of isotropic transmission.} This corroborates the practical significance of having a good quality LSE for realizing the array gains.
\subsection{Impact of Key Design Parameters on  Throughput Fairness}\label{sec:key-insights}
Here we quantify the impact of key system  parameters on the achievable fair BSC throughput $\mathrm{R_J}$ as returned by the proposed Algorithm~\ref{Algo:J-Opt}. Specifically, we consider the variation of four parameters: (i) field size $L$, varying from $20$m to $200$m, (ii) antenna array size ranging from $4$  to $20$ for $\mathcal{R}$, (iii) number of tags  $M$ varying between $1$ and $12$, (iv) effective backscattered SNR $\overline{\gamma}\triangleq\frac{p_t\overline{\beta}^2}{\sigma_{\mathrm{w}_{\mathcal{R}}}^2}$ ranging from $0$dB to $40$dB. The corresponding results plotted in Fig.~\ref{fig:Norm_R} show that $\mathrm{R_J}$ gets respectively enhanced by $6.4$dB and $78.0$dB when $N$ increases from $4$ to $20$ and $\overline{\gamma}$ improves from $0$dB to $40$dB. Similarly, with the increase in $M$ from $1$ to $12$ and $L$ from $20$m to $200$m, the underlying $\mathrm{R_J}$ gets degraded by $48.5$dB and $27.7$dB, respectively. These results are critical because they can be used in the smart designing of key parameters like $N,L,M,\overline{\gamma}$ in a MIMO reader-assisted multi-tag BSC setting so as to achieve a desired fair throughput $\mathrm{R_J}$ performance.

\begin{figure}[!t]	
		\centering\includegraphics[width=3.5in]{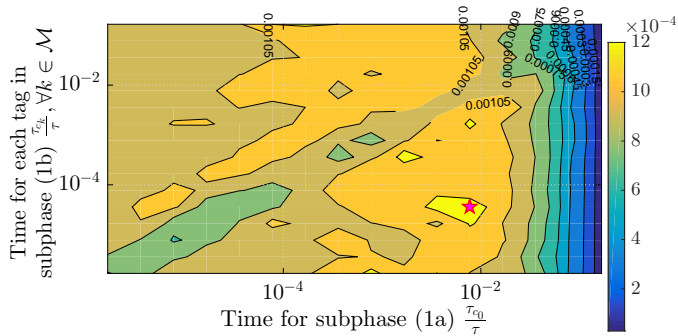}		
		\caption{Optimal max-min rate $\mathrm{R_J}$ as returned by Algorithm~\ref{Algo:J-Opt} for different values of the CE times $\tau_{c_0}$ and $\tau_{c_k},\forall k\in\mathcal{M}$.}
		\label{fig:Tc_2D}\color{black} 
\end{figure}

Next we shift focus to another key parameter, CE time $\tau_c$, that actually is critical in maintaining the optimal tradeoff between CE quality and spectral efficiency. {Here, we would like to first lay stress upon the fact that the analytical CE and ID time optimization requires strong prior knowledge of the channel statistics for both the backscattered links and UAR, which are difficult to obtain in practise. Also, even if we assume that those statistics are available, the underlying ergodic capacity or average throughput expression for each tag would be hard to derive in closed-form. Furthermore, as the underlying problem is nonconvex, we have restricted ourselves to numerical investigation of optimal time allocation. Therefore, noting that the CE time has two subphases, (1a) and (1b), as discussed in Section~\ref{sec:Two-Phase-CE}, we present a contour plot in Fig.~\ref{fig:Tc_2D} to show the variation of the optimal max-min rate $\mathrm{R_J}$  returned by Algorithm~\ref{Algo:J-Opt} for different values of $\tau_{c_0}$ and $\tau_{c_k}=\frac{\tau-\tau_{c_0}}{M},\forall k\in\mathcal{M}.$ We have also plotted the jointly-optimal CE time allocation $\left(\tau_{c_0},\tau_{c_k}\right)$ via the magenta-coloured starred-point in Fig.~\ref{fig:Tc_2D}. Here, we notice that for a given time allocation $\sum_{k=1}^{\mathcal{M}}\tau_{c_k}$ for the subphase (1b), the optimal time allocation $\tau_{c_0}$ for the subphase (1a) decreases with the increasing value of then former. In other words, the optimal $\tau_{c_0}$ is relatively lower for higher value of $\tau_{c_k}s$. Since, $\mathrm{R_J}$ changes very slowly with the variation of $\tau_{c_k}$, we notice that the latter has lesser significant impact. In contrast, $\tau_{c_0}$ is more critical because $\mathrm{R_J}$ is much more sensitive to the changes in the time allocated to sub-phase (1a) for UAR estimation and compensation. Also, in general, the value of $\tau_{c_0}$ should be less than ${0.01}{\tau}$. Due to this, we have set $\tau_{c_k}=\frac{\tau}{10(M+1)},\forall k\in\mathcal{M}\cup\{0\},$ as the default value for CE time parameters.} 

\begin{figure}[!t] 	
		\centering\includegraphics[width=3.4in]{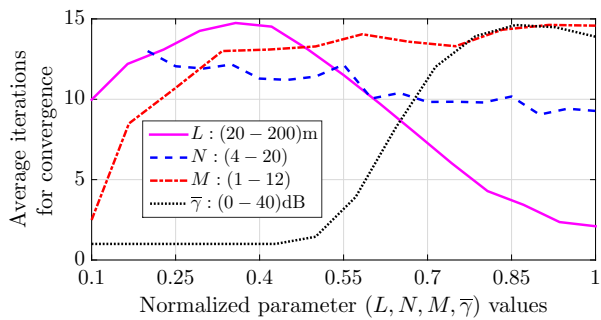}		
		\caption{Average number of iterations required by Algorithm~\ref{Algo:J-Opt} for achieving the acceptable tolerance $\epsilon$ for $\sigma_{\mathcal{R}}$ at  termination.}
		\label{fig:Norm_conv}  
\end{figure}

\subsection{Verifying the Convergence Claims of Proposed Algorithm}\label{sec:verify-conv}
In this section, we corroborate the practical utility of Algorithm~\ref{Algo:J-Opt} by verifying the fast convergence claims. Considering the same four parameters as in Section~\ref{sec:key-insights}, via Fig.~\ref{fig:Norm_conv} we show the average number of iterations required for obtaining the optimal max-min rate $\mathrm{R_J}$ (as plotted in Fig.~\ref{fig:Norm_R}). The average number of iterations for the different values of $L,N,M,$ and $\overline{\gamma}$ are $9,10, 12,$ and $6$, respectively. Also, comparing Figs.~\ref{fig:Norm_R} and~\ref{fig:Norm_conv}, we notice a correlation between the optimal max-min rate $\mathrm{R_J}$ and the number of iterations needed in achieving it. The general trend noted is that larger number of iterations are needed in the scenarios where higher $\mathrm{R_J}$ can be achieved.  More importantly, in most  cases, fewer than ten iterations are needed for realizing acceptable performance (where the standard deviation among rates of the tags is less than $\sigma_{\mathrm{R}}$).

\begin{figure}[!t]	
		\centering\includegraphics[width=3.48in]{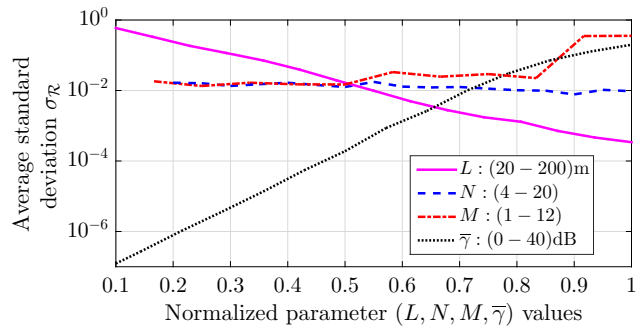}		
		\caption{Variation of the average deviation $\sigma_{\mathcal{R}}$ at  termination of Algorithm~\ref{Algo:J-Opt} for different system parameters like $L,N,M,\overline{\gamma}$.}
		\label{fig:Norm_std}    
\end{figure} 
Other than the iteration count, we also plot the value of $\sigma_{\mathrm{R}}$ at the termination of Algorithm~\ref{Algo:J-Opt} in Fig.~\ref{fig:Norm_std}. The average values of $\sigma_{\mathrm{R}}$ for varying $L,N,M,$ and $\overline{\gamma}$ values are $0.09,0.01, 0.07,$ and $0.03$, respectively. This implies at the termination of Algorithm~\ref{Algo:J-Opt}, the optimal TRX returned satisfies the optimality condition for CTM as stated in Lemma~\ref{lem:equal-R}. Thus, Figs.~\ref{fig:Norm_conv} and~\ref{fig:Norm_std} combinedly verify the fast convergence claims made for the proposed solution methodology in Section~\ref{sec:low}. 

\subsection{Comparison Against  Relevant Benchmark Scheme}\label{sec:comp}
In this part, we conduct a performance comparison study to corroborate the significance of proposed designs over the relevant benchmark scheme~\cite{WPCN-MIMO-CTM} that investigated the  CTM for WPCN. This benchmark adopted ZF as the detector design and precoder was set as below~\cite[eq. (8)]{WPCN-MIMO-CTM}
\begin{eqnarray}
\mathbf{f}_{\rm B}\triangleq\sqrt{p_t}\,\sum_{k=1}^{M}\sqrt{\frac{\frac{1}{\beta_k^2}}{\sum_{i=1}^{M}\frac{1}{\beta_i^2}}}\;\frac{\widehat{\mathbf{h}}_k^*}{\norm{\widehat{\mathbf{h}}_k^*}}.
\end{eqnarray}
\begin{figure}[!t] 
		\centering\includegraphics[width=3.48in]{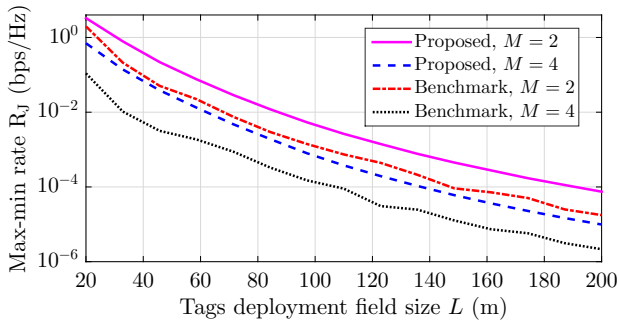}		
		\caption{Comparison of max-min rate as achieved by  proposed and benchmark TRX designs for different values of $L$ and $M$.}\label{fig:L_M}  
\end{figure}

We considered above as  benchmark because~\cite{WPCN-MIMO-CTM}  investigated maximizing the minimum throughput among EH users in WPCN, and as mentioned in Section~\ref{sec:rw} this set-up involving product channels has similar rate definition to as in BSC. Also, as far as BSC is concerned, this is the first work investigating TRX designing for CTM. For comparison, we first plot the variation of the max-min rate $\mathrm{R_J}$ as returned by  Algorithm~\ref{Algo:J-Opt} and benchmark scheme for different values of $L$ and $M$ in Fig.~\ref{fig:L_M}. We observe that with the field size  increased from $20$m to $200$m, $\mathrm{R_J}$ reduces by $46.5$dB for $M=2$ and by $48.5$dB for $M=4$. Therefore, the field size $L$ should be smartly selected based on the underlying tags density. Here, for $M=2$, the achievable max-min rate using the TRX design returned by Algorithm~\ref{Algo:J-Opt} is $6.7$ times more than that for $M=4$. Whereas, the proposed design provides  $3.8$ and $6.2$ times more max-min rate than benchmark for $M=2$ and $M=4$, respectively.

\begin{figure}[!t]	
		\centering\includegraphics[width=3.48in]{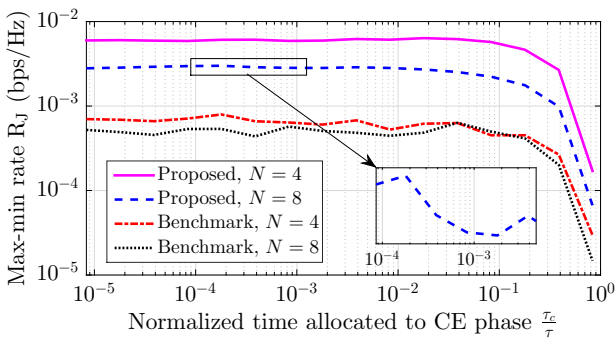}		
		\caption{Variation of the max-min rate for the proposed and benchmark designs for different CE time and array size values.}\label{fig:Tc_N}   
\end{figure} 
Next we  plot the max-min rate $\mathrm{R_J}$ with the proposed and  benchmark designs for varying values of CE time $\tau_c$ in Fig.~\ref{fig:Tc_N} for $N=4$ and $N=8$. Here, the irregularity in the trend followed by the max-min rate, especially for the benchmark scheme, is due to its non-concavity in $\tau_c$. {In fact even after following the numerical insights from Fig.~\ref{fig:Tc_2D} and setting the time for subphases (1a) and (1b) to be equal to $\tau_c$, we observe from Fig.~\ref{fig:Tc_N} that the underlying objective max-min rate $\mathrm{R_J}$ is still non-concave in $\tau_c$.}  We notice that the optimal $\tau_c$ should be less than $\frac{\tau}{100}$ for any array size $N$. Further, as the array size increases from  $N=4$ to $N=8$, the achievable max-min rate using the proposed design gets enhanced by $2.3$ times of its value at $N=4$. Moreover, there are significantly high gains of  about $5.4$-fold and $9.4$-fold increase in the fair BSC throughput as achieved by the proposed design over the benchmark one for $N=4$ and $N=8$, respectively. 
 
\begin{figure}[!t] 
		\centering\includegraphics[width=3.48in]{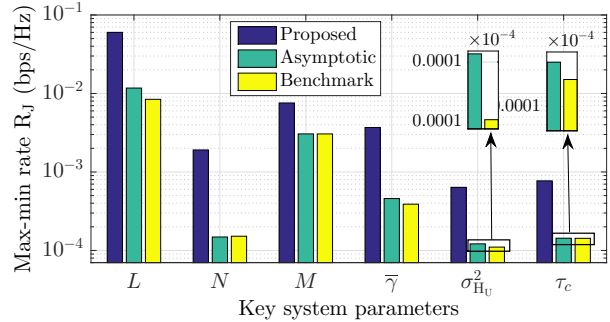}		
		\caption{Performance comparison of the proposed TRX design against the asymptotically-optimal and benchmark ones.}
		\label{fig:comp} 
\end{figure}
Finally to summarize, we plot the average max-min BSC rate as achieved by the proposed TRX design  obtained via Algorithm~\ref{Algo:J-Opt} and the asymptotically-optimal ones as obtained using Algorithm~\ref{Algo:Asy-Opt} against the benchmark in Fig.~\ref{fig:comp} over different values of critical system parameters $L,N,M,\overline{\gamma},\sigma_{\rm H_U}^2,$ and $\tau_c$. {Here, the variation of the former four parameters is same as mentioned in Section~\ref{sec:key-insights}, whereas the latter two are varied as: $\sigma_{\rm H_U}^2$ from $-90$dBm to $-170$dBm to incorporate relative strength of UAR that influences the CE errors, and with $\tau_c=\sum_{i=0}^{M}\tau_{c_i}$ where $\tau_{c_i}$ is varied between $\frac{10^{-5}\tau}{M+2}$ and $\frac{\tau}{M+2},\forall i\in\mathcal{M}\cup\{0\}$.} For the asymptotic case, we have plotted the one that leads to the higher max-min rate between the TRX designed for the low and high SNR regimes. We observe that the average gain provided by the proposed asymptotically-optimal designs over the benchmark scheme is about $11\%$ and this gap is more significantly visible for low SNR regimes as represented via variation of $L$ and $\overline{\gamma}$.  Overall, on an average terms, the proposed design $\left(\mathbf{f}_{\rm J},\mathbf{G}_{\rm J}\right)$ provides more than $6.5$-fold and $7.2$-fold enhancement in the average max-min rate as respectively achieved by the asymptotically-optimal and  benchmark schemes.  The main reason for such high gains is that the proposed TRX designs are specifically designed for BSC settings keeping in mind the underlying product channel based throughput definitions and resource-constraints for the tags. Also, the significant gains over the asymptotically-optimal designs highlights the relative importance of the TRX design returned by Algorithm~\ref{Algo:J-Opt} over the ones obtained via Algorithm~\ref{Algo:Asy-Opt}.  This improvement in fair backscattered throughput (or the max-min rate) gets even more enhanced for larger array sizes as shown via the result for variation of $N$ in Fig.~\ref{fig:comp}. Thus, we realize that the proposed CE protocol along with the optimal TRX designs $\left(\mathbf{f}_{\rm J},\mathbf{G}_{\rm J}\right)$ and $\left(\mathbf{f}_{\rm A},\mathbf{G}_{\rm A}\right),\forall\rm{A}\in\left\lbrace L,H\right\rbrace$, clearly outperforming existing design over the entire key system parameter ranges, can help in achieving significant enhancement in terms of CTM.

\section{Concluding Remarks}\label{sec:concl}  
This work presented a novel CE algorithm involving optimal TRX designing at the full-duplex MIMO reader for throughout fairness maximization in a multi-tag BSC setting. Since, the underlying CTM problem was nonconvex, an efficient low-complexity iterative algorithm was proposed, which outperformed the existing TRX designs. The key merits of the proposed LS-based CE algorithm is that it does not require any prior information and also takes care of both environmental UAR and the resource-constraints  of  tags in assisting the ongoing estimation at the reader.  To gain practical insights on the optimal precoder and detector at the multiantenna reader, both individually and asymptotically TRX designs were also discoursed. For validating the quality of the proposed LSE and the optimal TRX design, extensive numerical simulations were conducted. We have shown that the proposed novel designs yield a significant seven times higher fair backscattered throughput as compared to the ones achieved by the existing benchmarks. As shown via numerical investigation, these gains are strongly influenced by the CE quality and key parameters like reader's  array size, tags count, and field size.  

In future, this baseline work can be extended to investigate the optimal performance of other advanced BSC settings like ambient, bi-static, or the ones with multi-antenna tags, so as to meet the timely practical sustainability demand of low-power wireless devices for IoT applications. {We would also like to derive the analytical expressions for optimal CE time allocation and design the underlying pilot sequences so as to maximize the achievable backscattered throughput while considering the impedance-switching based tags' signal modulation and  constellations in mind.}

\bibliographystyle{IEEEtran} 


\end{document}